\documentclass[final]{IEEEtran}
\usepackage[latin9]{inputenc}
\usepackage{color}
\usepackage{units}
\usepackage{amsthm}
\usepackage{amsmath}
\usepackage{amssymb}
\usepackage{graphicx}
\usepackage{url}
\usepackage{cite}
\usepackage{epsfig}
\usepackage{algorithm}
\usepackage{algorithmic}
\usepackage{comment}
\usepackage{color}

\DeclareMathOperator*{\minimize}{minimize}

\DeclareMathOperator*{\st}{subject\;to}

\makeatletter

\theoremstyle{plain}

\theoremstyle{plain}
\newtheorem{prop}{\protect\propositionname}
\theoremstyle{plain}
\newtheorem{cor}{\protect\corollaryname}
\theoremstyle{plain}

\theoremstyle{remark}
\newtheorem{rem}[]{\protect\remarkname}

\ifCLASSOPTIONcompsoc
\usepackage[caption=false,font=normalsize,labelfont=sf,textfont=sf,labelformat=simple]{subfig}
\else
\usepackage[caption=false,font=footnotesize,labelformat=simple]{subfig}
\fi

\makeatother

\providecommand{\remarkname}{Remark}
\providecommand{\corollaryname}{Corollary}
\providecommand{\lemmaname}{Lemma}
\providecommand{\propositionname}{Proposition}
\providecommand{\theoremname}{Theorem}

\begin{document}

\title{Secure Video Streaming in Heterogeneous Small Cell Networks with Untrusted Cache Helpers}

\author{{Lin~Xiang,~\IEEEmembership{Student~Member,~IEEE,} Derrick~Wing~Kwan~Ng,~\IEEEmembership{Senior~Member,~IEEE,} \\ Robert~Schober,~\IEEEmembership{Fellow,~IEEE,} and Vincent~W.S.~Wong,~\IEEEmembership{Fellow,~IEEE}}

\thanks{The work of D. W. K. Ng was supported under Australian Research Council's Discovery Early Career Researcher Award funding scheme (DE170100137). The work of R. Schober was supported by the Alexander von Humboldt Professorship Program. The work of V.W.S. Wong was supported by the Natural Sciences and Engineering Research Council of Canada. Part of this work has been accepted for presentation at the \emph{IEEE Global Commun. Conf. (Globecom)}, Singapore, Dec. 2017 \cite{Xiang17:SVC}. 
}
\thanks{L.~Xiang and R.~Schober are with the Institute for Digital Communications, Friedrich-Alexander University of Erlangen-Nuremberg, Erlangen 91058, Germany (Email: \{lin.xiang, robert.schober\}@fau.de).}  
\thanks{D. W. K.~Ng is with the School of Electrical Engineering and Telecommunications, University of New South Wales, Sydney, NSW 2052, Australia (Email: w.k.ng@unsw.edu.au).}
\thanks{V.W.S.~Wong is with the Department of Electrical and Computer Engineering, University of British Columbia, Vancouver, BC V6T 1Z4, Canada (Email: vincentw@ece.ubc.ca).}
\vspace{-.2cm}
}

\maketitle

\vspace{-1cm}
\begin{abstract}
This paper studies secure video streaming in cache-enabled small cell networks, where some of the cache-enabled small cell base stations (BSs) helping in video delivery are \emph{untrusted}. Unfavorably, caching improves the eavesdropping capability of these untrusted helpers as they may intercept both the cached and the delivered video files. To address this issue, we propose joint caching and scalable video coding (SVC) of video files to enable secure cooperative multiple-input multiple-output (MIMO) transmission and, at the same time, exploit the cache memory of both the trusted and untrusted BSs for improving the system performance. Considering imperfect channel state information (CSI) at the transmitters, we formulate a two-timescale non-convex mixed-integer robust optimization problem to minimize the total transmit power required for guaranteeing the quality of service (QoS) and secrecy during video streaming. We develop an iterative algorithm based on a modified generalized Benders decomposition (GBD) to solve the problem optimally, where the caching and the cooperative transmission policies are determined via offline (long-timescale) and online (short-timescale) optimization, respectively. Furthermore, inspired by the optimal algorithm, a low-complexity suboptimal algorithm based on a greedy heuristic is proposed. Simulation results show that the proposed schemes achieve significant gains in power efficiency and secrecy performance compared to several baseline schemes. 
\end{abstract}


\section{Introduction}

\IEEEPARstart{S}{mall} cells are among the most promising solutions for meeting the enormous capacity requirements introduced by video streaming applications in the fifth-generation (5G) wireless networks \cite{Schober5G}. By densely deploying low-power base stations (BSs), both the spectral and energy efficiencies of wireless communication systems can be improved significantly. However, to achieve these performance gains, high-capacity secure backhaul links are required to transport the video files from the Internet to the small cell BSs. While wireless backhauling is usually preferred for small cells due to its low cost and high flexibility in deployment \cite{Ge14Network:Backhaul}, the capacity provided by wireless backhauling is often insufficient, which deteriorates the overall system performance \cite{Rost14MCOM:CloudRAN} and limits the maximum number of concurrent streaming users/connections. Moreover, since wireless transmission is susceptible to eavesdropping, the security of wireless backhauling is a fundamental concern for 5G wireless networks.

Recently, wireless caching has been proposed as a viable solution to enhance the capacity of small cell backhauling \cite{Paschos16WC,Chen:JSAC17:UAV,Caire13IT:FemtoCaching,LiuJSAC16:EE,Liu14TSP:CoMP,Xiang17TVT:CLCaching,Maddah14:Centralized,Maddah15:Decentralized,Zhang15:Coded,Naderializadeh17,Zewail:ISIT17}. Built upon the content-centric networking paradigm, in wireless caching, the most popular contents are pre-stored at the access points or BSs in close proximity of the user equipments (UEs). Consequently, the backhaul traffic is offloaded by reusing the cached content \cite{Paschos16WC,Chen:JSAC17:UAV}. Caching as an alternative to small cell backhauling was first investigated in \cite{Caire13IT:FemtoCaching}, where caching was shown to also substantially reduce the average downloading delay. Besides, caching improves the energy efficiency of wireless backhauling systems as was shown in \cite{LiuJSAC16:EE}. In \cite{Liu14TSP:CoMP}, caching was optimized to facilitate power-efficient cooperative multiple-input multiple-output (MIMO) transmission in small cell networks. In \cite{Xiang17TVT:CLCaching}, joint caching and buffering for small cell networks was proposed to overcome the backhaul capacity bottleneck and the half-duplex transmission constraint simultaneously to enable fast downloading of video files. In \cite{Maddah14:Centralized,Maddah15:Decentralized}, coded caching was introduced, which reduces the backhaul load by exploiting coded multicast transmission for simultaneous delivery of different files. Coded caching was extended to various network scenarios, see \cite{Zhang15:Coded,Naderializadeh17,Zewail:ISIT17} and references therein.

On the other hand, although communication secrecy is of high importance in wireless networks, providing security for networks employing wireless caching has been a major challenge. This is because current video streaming applications, e.g. YouTube and Netflix, mainly rely on end-to-end encryption schemes such as the hypertext transfer protocol secure (HTTPS) \cite{Tescorla00https} to ensure communication security. However, with such schemes, the benefits of content-centric networking vanish as encrypted contents are uniquely defined for each user request and cannot be reused to serve other user requests \cite{Paschos16WC}. For this reason, caching was mainly considered for content without security restrictions in the literature \cite{Paschos16WC,Chen:JSAC17:UAV,Caire13IT:FemtoCaching,LiuJSAC16:EE,Liu14TSP:CoMP,Xiang17TVT:CLCaching}. To overcome this limitation, physical layer security (PLS) schemes for wireless caching were proposed in \cite{Xiang17TWC:SecCaching,LandICC16:HetNets,Zewail:CNS16}. As PLS techniques rely on wiretap channel coding instead of source encryption, content can still be reused at the wireless edge for secure wireless transmission, and hence, caching and PLS are compatible. In \cite{Xiang17TWC:SecCaching}, cache-enabled cooperative MIMO transmission was shown to be an effective physical layer mechanism for increasing the secrecy rate for video delivery in homogeneous cellular networks. However, a secure backhaul for cache placement was required in \cite{Xiang17TWC:SecCaching}, which cannot always be guaranteed with wireless backhauling in practice. Considering an insecure backhaul, a secure cache placement strategy for heterogeneous cellular networks (HetNets) was developed in \cite{LandICC16:HetNets}, whereby eavesdroppers tapping the insecure backhaul can be prevented from obtaining a sufficient number of coded packets for successful recovery of the video file. Assuming caching at the end users, the authors of \cite{Zewail:CNS16} proposed a secure coded multicast scheme for relay networks to prevent end users and external eavesdroppers from intercepting the non-requested and the delivered files, respectively.

However, \cite{Xiang17TWC:SecCaching,LandICC16:HetNets,Zewail:CNS16} optimistically assumed that the (cache) helpers can be \emph{trusted} for cooperation and that the cache cannot be exploited for eavesdropping purposes. These assumptions may be unrealistic for HetNets. In particular, due to the distributed network architecture, cache-enabled small cell BSs can be \emph{untrusted} helpers, i.e., they may be potential eavesdroppers\footnote{In this paper, we only consider passive eavesdroppers which remain silent during eavesdropping. Studying the case of active eavesdroppers such as jamming and spoofing attackers \cite{Xiao16Access} is an interesting topic for future work.}, and hence, may not cooperate altruistically \cite{Wright15:Hacking,Mantas15:Security:5G,3GPP:TR33820}. Examples of untrusted helpers include home-owned and open-access small cell BSs which can be easily manipulated by third parties to eavesdrop premium video streaming services, for which they have not paid, and/or users' private video files. In contrast to trusted small cell BSs deployed and owned by the service provider, at these untrusted small cell BSs, user data is left unprotected and prone to eavesdropping because the small cell BS itself is responsible for encrypting and decrypting the user data before forwarding it to the macro BS and the intended users, respectively \cite{Wright15:Hacking,Mantas15:Security:5G,3GPP:TR33820}. Moreover, different from the case of cache-disabled eavesdroppers%
\footnote{The case considered in this paper is more general than that in \cite{LandICC16:HetNets,Xiang17TWC:SecCaching}. In fact, the eavesdroppers in \cite{LandICC16:HetNets,Xiang17TWC:SecCaching} can be viewed as untrusted helpers with zero cache capacity. %
} considered in \cite{LandICC16:HetNets,Xiang17TWC:SecCaching}, the cache memory equipped at the untrusted helpers unfavorably enhances their eavesdropping capability as they can intercept both the cached and the delivered video data, and utilize the cached video data as side information to improve reception.

Two fundamental questions need to be addressed when cache helpers are untrusted: (a) \emph{Can cooperation with untrusted helpers still yield secrecy benefits? That is, can the cache deployed at the untrusted helpers be utilized to improve the system performance?} If so, (b) \emph{how to cache and cooperate intelligently to reap the possible performance gains?} To our knowledge, state-of-the-art small cell networks perform only passive authentication of BSs and completely exclude untrusted BSs from participating in cooperative transmission~\cite{Mantas15:Security:5G,3GPP:TR33820}. However, in this case, untrusted BS cannot be used to improve the system performance\footnote{We note that, as untrusted BSs present the man-in-the-middle threat to wireless networks, HTTPS also cannot facilitate secure cooperative transmission \cite{Callegati09man}.}.

In \cite{He:IT10}, untrusted helpers without caching have been investigated for relay networks. It was shown that cooperation with compress-and-forward relays yields a positive secrecy rate even if the relays are untrusted. However, the problem studied in this paper is more challenging as the untrusted helpers can cache content to enhance their eavesdropping capability. Thus, the solutions proposed in \cite{He:IT10} are not applicable and a new study is needed. In this paper, to facilitate secure cooperative transmission with untrusted cache helpers, we propose an advanced caching scheme that combines scalable video coding (SVC) and cooperative MIMO transmission. Specifically, each video file is encoded by SVC into base-layer subfiles, containing basic-quality and independently decodable video information, and enhancement-layer subfiles, containing high-quality video information which is decodable only after the base layer has been successfully decoded. By caching the enhancement-layer subfiles across all BSs and the base-layer subfiles only across trusted BSs, secure cooperation of all BSs is enabled by exploiting the encoding and decoding structure of SVC. Thereby, the large virtual transmit antenna array formed by all BSs that have cached the same subfile introduces additional degrees of freedom (DoFs) which may be utilized for secure and power-efficient video streaming.

To reap the cache-enabled secrecy benefits, a centralized framework for caching and delivery optimization is adopted. Hence, the proposed architecture follows the cloud radio access network (CRAN) philosophy \cite{checko2015cloud,Rost14MCOM:CloudRAN} which has been advocated for next-generation HetNets for cooperative MIMO transmission \cite{Zhao13TWC:BchOpt,Zhuang14TSP:AsyCoop} and advanced resource allocation \cite{Ng15TWC}. In the conference version of this paper \cite{Xiang17:SVC}, we investigated cache-enabled secure transmission by assuming perfect knowledge of all channels. However, in practice, the channel state information (CSI) gathered at the central controller, e.g. the macro BS, is imperfect due to quantization noise and feedback delay, which deteriorates the system performance and has to be taken into account for system design. To mitigate the information leakage from the trusted BSs to the untrusted BSs, artificial noise (AN) based jamming is applied in this paper. In the literature \cite{Negi08TWC:AN,LiTSP11:Opt-Robust,Ng15TWC}, AN has been employed to effectively reduce the receive signal-to-interference-plus-noise ratio (SINR) at the eavesdropper without interfering the desired users. In this paper, we consider cooperative AN transmission by the trusted BSs for power-efficient jamming to combat the eavesdropping of the untrusted BSs. By considering untrusted BSs and imperfect CSI, we jointly optimize caching and cooperative data and AN transmission for a secure and power-efficient system design. In particular, a two-timescale robust optimization problem is formulated for minimization of the transmit power required for secure video streaming under imperfect CSI. The main contributions of this paper can be summarized as follows: 
\begin{itemize}
\item We study a new secrecy threat in small cell networks originating from untrusted cache helpers, i.e., cache-enabled eavesdropping small cell BSs. To facilitate secure cooperative MIMO transmission of trusted and untrusted small cell BSs, we propose a secure caching scheme based on SVC.

\item We optimize the caching and the cooperative delivery policies for minimization of the transmit power while guaranteeing quality-of-service (QoS) and communication secrecy under imperfect CSI. We show that the optimal delivery decisions can be obtained by semidefinite programming (SDP) relaxation with probability one under mild conditions. For the optimal caching decisions, an optimal iterative algorithm is developed based on a modified generalized Benders decomposition (GBD). To reduce the computational complexity, a polynomial-time suboptimal greedy scheme is also proposed. 

\item Our simulation results show that the proposed robust schemes can efficiently exploit the cache capacities of both trusted and untrusted small cell BSs to enable power-efficient and secure video streaming in heterogeneous small cell networks. 
\end{itemize}

\indent The remainder of this paper is organized as follows. In Section~\ref{sec:System-Model}, we present the system model for cooperative secure video streaming in the presence of untrusted cache helpers. The formulation and solution of the proposed optimization problem are provided in Sections~\ref{sec:Problem-Formul} and~\ref{sec:Problem-Sol}, respectively. In Section~\ref{sec:Simulation-Results}, we evaluate the performance of the proposed schemes and compare it to that of several baseline schemes. Finally, Section~\ref{sec:Conclusion} concludes the paper.

\emph{Notations:} $\mathbb{R}$ and $\mathbb{C}$ denote the sets of real and complex numbers, respectively; $\Re\{z\}$ denotes the real part of $z \in \mathbb{C}$; $\mathbf{I}_{L}$ is an $L\times L$ identity matrix; $\mathbf{1}_{M\times N}$ and $\mathbf{0}_{M\times N}$ are $M \times N$ all-one and all-zero matrices, respectively; $(\cdot)^{T}$ and $(\cdot)^{H}$ are the transpose and complex conjugate transpose operators, respectively; $\left\Vert \cdot\right\Vert _{\ell}$ denotes the $\ell$-norm of a vector; $\left\Vert \cdot\right\Vert _{F}$, $\mathrm{\mathrm{tr}}(\cdot)$, $\mathrm{rank}(\cdot)$, $\det(\cdot)$, and $\lambda_{\max}(\cdot)$ denote the Frobenius norm, the trace, the rank, the determinant, and the maximal eigenvalue of a square matrix, respectively; $\mathbb{E}(\cdot)$ is the expectation operator; the circularly symmetric complex Gaussian distribution is denoted by $\mathcal{CN} (\boldsymbol{\mu}, \mathbf{C})$ with mean vector $\boldsymbol{\mu}$ and covariance matrix $\mathbf{C}$; $\sim$ stands for ``distributed as"; $\mathrm{diag}(\mathbf{x})$ is a diagonal matrix with the diagonal elements given by vector $\mathbf{x}$; $|\mathcal{X}|$ represents the cardinality of set $\mathcal{X}$; $\mathbf{A}\succeq\mathbf{0}$ and $\mathbf{A}\succ\mathbf{0}$ indicate that matrix $\mathbf{A}$ is positive semidefinite and positive definite, respectively; finally, $\left[x\right]^{+}$ stands for $\max\{0,x\}$.

\section{\label{sec:System-Model}System Model}

\subsection{Network Topology}

\begin{table*}
\centering
\protect\protect\caption{\label{tab2}List of Key Notations.}
\small{ 
\begin{tabular}{|c|c|}
\hline 
 {$\mathcal{M}$, $\mathcal{M_T}$, $\mathcal{M_U}$}    &  {Sets of $M$ BSs, $J$ trusted BSs, and $M-J$ untrusted BSs } \tabularnewline
\hline
{$N$, $N_{\mathcal{T}}$, $N_{\mathcal{U}}$} & {Total number of antennas at all BSs, trusted BSs, and untrusted BSs}
\tabularnewline
\hline 
{$\mathcal{M}_{f,l}^{\mathrm{Coop}}$}  & {Subset of cooperating BSs for delivery of subfile $(f,l)$}
\tabularnewline
\hline 
 {$\mathcal{K}$, $\mathcal{F}$, $\mathcal{L}$, $\mathcal{T}_0$} & {Sets of $K$ users, $F$ video files, $L$ layer subfiles per file, and $T_0$ time slots}\tabularnewline
\hline 
{$\boldsymbol{\rho}\triangleq(k,f)$, $\mathcal{S}$} & {Request of user $k$ for file $f$ and set of user requests }\tabularnewline
\hline 
{$\kappa(\boldsymbol{\rho})$, $f(\boldsymbol{\rho})$} & Requesting UE and requested file corresponding to $\boldsymbol{\rho}$ 
\tabularnewline
\hline 
{$q_{f,l,m} \in \{0,1\}$} & {Binary cooperative delivery decisions for subfile $(f,l)$ at BS $m$}\tabularnewline
\hline
{$s_{\boldsymbol{\rho},l,t}$} & {Source symbol of subfile $(f,l)$ for serving request $\boldsymbol{\rho}$ at time $t$}
\tabularnewline
\hline 
{$\mathbf{w}_{\boldsymbol{\rho},l,m,t}$, $\mathbf{w}_{\boldsymbol{\rho},l,t}$} & {Beamforming vectors of BS $m$ and BS set $\mathcal{M}$ for sending symbol $s_{\boldsymbol{\rho},l,t}$}\tabularnewline
\hline 
{$\mathbf{v}_t$, $\mathbf{V}_t$ } & { AN and its covariance matrix at time $t$}\tabularnewline
\hline 
{$C_m^{\max}$ } & { Cache size at BS $m$ }\tabularnewline
\hline 
{$R_{\boldsymbol{\rho},l,t}$, $R_{\boldsymbol{\rho},l,t}^{\mathrm{sec}}$ } & {Achievable rate and achievable secrecy rate at user $\kappa(\boldsymbol{\rho})$ for decoding $s_{\boldsymbol{\rho},l,t}$ }\tabularnewline
\hline 
{$R_{j,\boldsymbol{\rho},l,t}$} &  {Capacity of untrusted BS $j$ for eavesdropping symbol $s_{\boldsymbol{\rho},l,t}$}
\tabularnewline
\hline 
\end{tabular} }
\end{table*}

\begin{figure}
\vspace{-.5cm}
\centering
\subfloat[] {\includegraphics[width=3.0in]{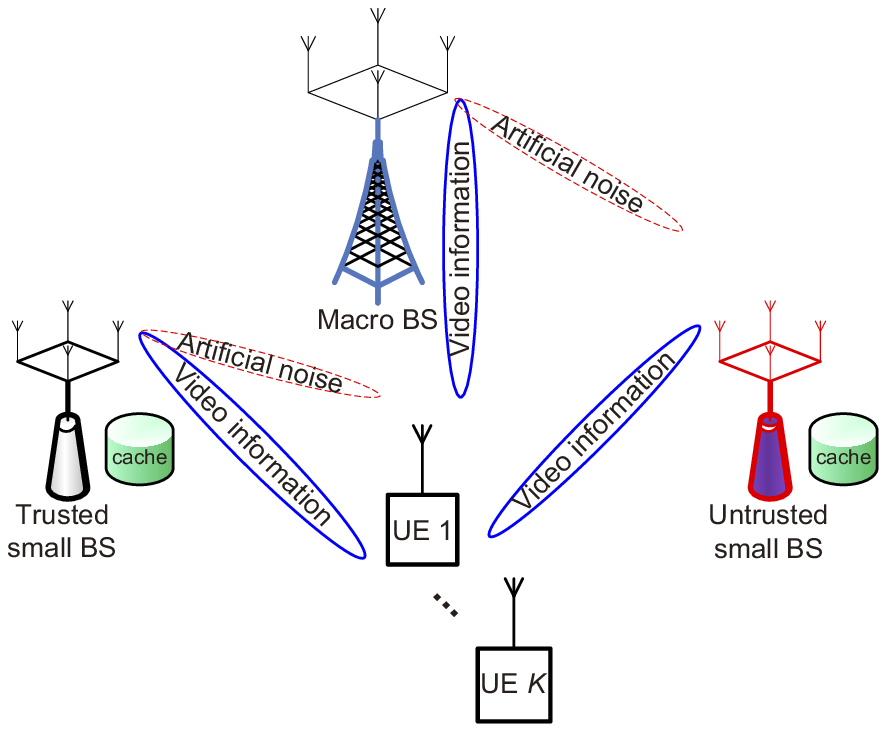} \label{fig1} } \quad 
\subfloat[] {\includegraphics[width=3.0in]{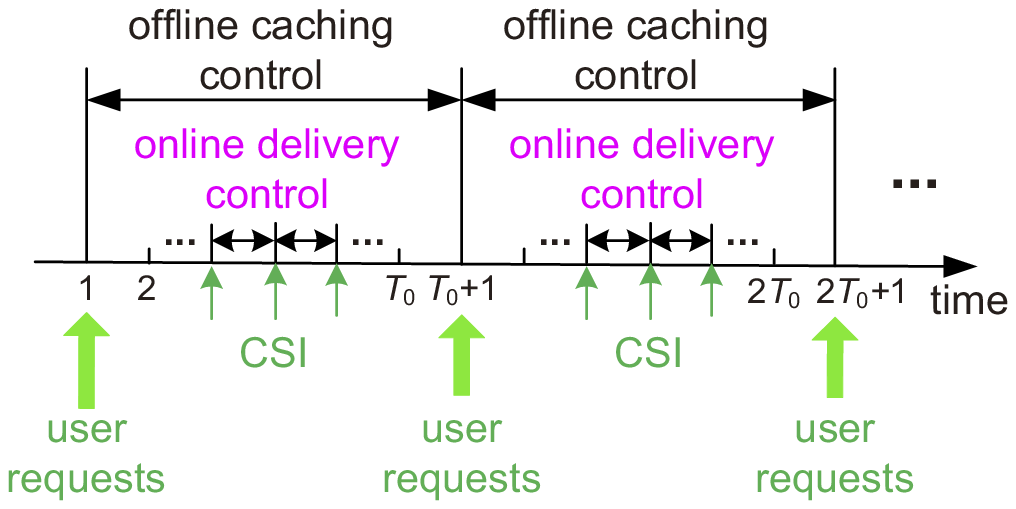} \label{fig1b} }
\vspace{-.1cm}
\protect\protect\caption{(a) System model for secure video delivery in a heterogeneous network, where a trusted and an untrusted small cell BS are distributed in the coverage area of a macro BS; (b) caching and delivery control are performed in two timescales.}
\end{figure}

We consider downlink wireless video streaming in a heterogeneous small cell network, where $M$ small cell BSs, each equipped with a cache memory of size $C_m^{\max}$ bits, are distributed in the coverage area of a macro BS, see Fig.~\ref{fig1}. For convenience, a list with key notations used in this paper is provided in Table~\ref{tab2}. Let $m\in\mathcal{M}\triangleq\left\{ 0,1,\ldots,M\right\} $ be the BS index, where $m=0$ refers to the macro BS. The macro BS is connected to the video server on the Internet via a dedicated secure high-capacity backhaul link such as optical fiber. For simplicity of notation, the backhaul to the macro BS is modeled as a cache with an equivalent capacity of $C_{0}^{\max}$ bits. In contrast, the small cell BSs are connected  to the macro BS via wireless backhaul links for convenience of deployment. Assume that $BS_{m}$, $m\in\mathcal{M}$, is equipped with $N_{m}$ antennas. The total number of transmit antennas is denoted by $N\triangleq\sum_{m\in\mathcal{M}}N_{m}$.

The video server owns a library of $F$ video files, indexed by $\mathcal{F}\triangleq\left\{ 1,\ldots,F\right\} $, to be streamed to $K$ single-antenna UEs, indexed by $\mathcal{K} \triangleq \left\{1,\ldots,K\right\}$. The size of file $f$ is $V_{f}$ bits. Employing SVC coding, as utilized e.g. for wireless video delivery in the H.264/Moving Picture Expert Group (MPEG)-4 standard \cite{Ohm05:SVC,Schwarz07:H264,Sanchez11idash}, each video file $f \in \mathcal{F}$ is encoded into one base-layer subfile, $\left(f,0\right)$, and $L-1$ enhancement-layer subfiles, $\left(f,l\right)$, $l\in\left\{ 1,\ldots,L-1\right\}$, where the information embedded in enhancement layer $l$ is used to refine the information contained in  the previous layers $0,\ldots,l-1$. Let $\mathcal{L}\triangleq\left\{ 0,\ldots,L-1\right\} $ be the index set of all layers. The size of subfile $\left(f,l\right)$ is $V_{f,l}$ bits. The base layer can be decoded independent of the enhancement layers. In contrast, enhancement layer $l\in \mathcal{L} \backslash \left\{0\right\} $ can be decoded only after layers $0,\ldots,l-1$ have already been decoded. Therefore, the layers have to be decoded in a sequential manner \cite{Ohm05:SVC}. Due to this specific encoding and decoding structure, only the base layer has to be protected in order to ensure communication secrecy. An eavesdropper, who cannot decode the base layer, will also not be able to decode any of the enhancement layers.

The small cell BSs serve as helpers of the macro BS in delivering the video files.  However, a subset of the small cell BSs are untrusted. These BSs may leak the cached video data and eavesdrop the transmitted video data while utilizing the cached data as side information. Let $\mathcal{M_{T}}\triangleq\left\{ 0,1,\ldots,J\right\} $ and $\mathcal{M_{U}}\triangleq\left\{ J+1,\ldots,M\right\} $ denote the sets of trusted and untrusted BSs having a total number of $N_{\mathcal{T}} \triangleq \sum_{m \in \mathcal{M}_{\mathcal{T}}} N_{m}$ and $N_{\mathcal{U}} \triangleq \sum_{m \in \mathcal{M}_{\mathcal{U}}} N_{m}$ antennas, respectively, where $J\le M$ and $N_{\mathcal{T}} + N_{\mathcal{U}} = N$. In this paper, we assume that the set of untrusted BSs, $\mathcal{M_U}$, is known. In practice, untrusted BSs may largely be home-owned and open-access small cell BSs which have insufficient security protection and can easily be compromised by third parties. Due to the eavesdropping and intensive processing, untrusted BSs may consume a large power and/or experience a long end-to-end latency even if the uplink and downlink throughputs are small. Hence, the power/delay versus throughput pattern of untrusted BSs is statistically different from that of trusted operator-owned BSs such that they constitute outliers. Therefore, by exploiting the power, delay, and throughput records of all BSs collected by the service providers, the set of untrusted BSs can be estimated by applying state-of-the-art outlier detection methods, e.g., supervised and unsupervised learning techniques \cite{Zhang10:Outlier,Gogoi11survey,Hodge04survey}.

The considered system is time slotted and the duration of a time slot is smaller 
than the channel coherence time. We consider a two-timescale control for caching and delivery. As shown in Fig.~\ref{fig1b}, the video files in the cache are updated every $T_{0}$ time slots, referred as one period, based on the historical profiles of user preferences and CSI. In contrast, the video delivery decisions are determined in each time slot based on the actual requests of the users and instantaneous CSI. We have $T_{0}\gg1$, as the users' preferences vary on a much slower scale (e.g., from day to day) than the user requests. For notational simplicity, we consider the system only during one typical period $\mathcal{T}_{0}\triangleq\left\{ 1,\ldots,T_{0}\right\} $ and the corresponding time slots are indexed by $t\in\mathcal{T}_{0}$.

\subsection{Secure Video Caching and Delivery}

As the cache helpers in set $\mathcal{M_{U}}$ are untrusted, only the enhancement layers are cached at BSs $m \in \mathcal{M_{U}}$. Hence, the cached subfiles cannot be used by the untrusted BSs to reconstruct the original video files as long as they do not have access to the base-layer subfiles. Meanwhile, BSs that have the same base-layer or enhancement-layer subfile cached, can employ cooperative transmission for power-efficient and secure delivery of the subfile to the UEs. On the other hand, video files that are uncached at the small cell BSs can be delivered only by the macro BS. Let $q_{f,l,m}=1$ indicate that subfile $\left(f,l\right)$ is cached at BS $m$, and $q_{f,l,m}=0$ otherwise. The cache placement has to satisfy the condition
\vspace{-.2cm}
\begin{alignat}{1}
\textrm{C1:}\; &  q_{f,l,m} \in\left\{ 0,1\right\} ,\;\forall(f,l)\in\mathcal{F}\times\mathcal{L},\,\forall m\in\mathcal{M},\;\;\textrm{and}  \nonumber \\ 
  & q_{f,0,m}=0,\;\forall f\in\mathcal{F},\, m\in\mathcal{M_{U}},  
\end{alignat}
and the capacity constraint
\vspace{-.1cm}
\begin{alignat}{1}
\textrm{C2:}\;\sum\limits _{(f,l)\in\mathcal{F}\times\mathcal{L}} & q_{f,l,m}V_{f,l}\le C_{m}^{\max},\; m\in\mathcal{M}.
\end{alignat}

During data delivery, the set of BSs cooperating for the delivery of subfile $(f,l)$ is denoted by $\mathcal{M}_{f,l}^{\mathrm{Coop}} \triangleq \left\{ m \in \mathcal{M} \mid q_{f,l,m}=1\right\} $. Assume that a UE requests one file but possibly multiple layers of the file at a time. We denote a request from user $\kappa$ for file $f$ by $\boldsymbol{\rho} \triangleq (\kappa,f)$ and the set of requests by $\mathcal{S} \subseteq \mathcal{K} \times \mathcal{F}$. For convenience, the requesting UE and the requested file corresponding to $\boldsymbol{\rho}$ are denoted by $\kappa(\boldsymbol{\rho})$ and $f(\boldsymbol{\rho})$, respectively. Moreover, user $\kappa(\boldsymbol{\rho})$ may request $L_{\boldsymbol{\rho}}$ layers, indexed by $\mathcal{L}_{\boldsymbol{\rho}} \triangleq \left\{ 0,1,\ldots,L_{\boldsymbol{\rho}}-1\right\} $.

We assume a frequency flat fading channel for video data transmission. As the worst case, we assume that the untrusted BSs are full-duplex, i.e., they can  simultaneously eavesdrop the video information intended for the UEs and participate in the cooperative delivery of the cached files. In time slot $t \in \mathcal{T}_0$, the self-interference \cite{bharadia13FD,DayTSP12:FD} at BS $j\in \mathcal{M_{U}}$ caused by simultaneous reception and transmission at the same frequency is denoted by $\mathbf{c}_{j,t}$. Let $\mathbf{x}_{t}\in\mathbb{C}^{N\times1}$ denote the joint transmit signal of BS set $\mathcal{M}$. The received signals at user $\kappa(\boldsymbol{\rho})$ and the untrusted BSs, denoted by $y_{\boldsymbol{\rho},t} \in \mathbb{C}$ and $\mathbf{y}_{\mathcal{U},j,t} \in \mathbb{C}^{N_{j}\times1}$, $j \in \mathcal{M_{U}}$, respectively, are given by 
\vspace{-.1cm}
\begin{equation}
 y_{\boldsymbol{\rho},t} =\mathbf{h}_{\boldsymbol{\rho},t}^{H}\mathbf{x}_{t}+z_{\boldsymbol{\rho},t} \; \textrm{and}\;  
\mathbf{y}_{\mathcal{U},j,t}  =\mathbf{G}_{j,t}^{H}\mathbf{x}_{t}+\mathbf{c}_{j,t}+\mathbf{z}_{j,t},  
\end{equation}
where $\mathbf{h}_{\boldsymbol{\rho},t} = [\mathbf{h}_{\boldsymbol{\rho},0,t}^{H}, \dots,\mathbf{h}_{\boldsymbol{\rho},M,t}^{H}]^{H} \in \mathbb{C}^{N\times1}$ and $\mathbf{G}_{j,t} = [\mathbf{G}_{j,0,t}^{H}, \dots, \mathbf{G}_{j,j-1,t}^{H},\mathbf{0}_{N_j\times N_{j}}^{H}, \!\mathbf{G}_{j,j+1,t}^{H}, \dots, \mathbf{G}_{j,M,t}^{H}]^{H}$ $\in\mathbb{C}^{N\times N_{j}}$ are the channel vectors/matrices from BS set $\mathcal{M}$ to user $\kappa(\boldsymbol{\rho})$ and BS $j$, respectively. $\mathbf{h}_{\boldsymbol{\rho},m,t} \in \mathbb{C}^{N_{m}\times1}$ and $\mathbf{G}_{j,m,t} \in \mathbb{C}^{N_{m} \times N_{j}}$ model the channels between BS $m\!\in\! \mathcal{M}$ and the respective receivers. The term $\mathbf{0}_{N_j\times N_{j}}^{H}$ in the definition of $\mathbf{G}_{j,t}$ accounts for the fact that the self-interference at BS $j$ is included in $\mathbf{c}_{j,t}$. Furthermore, $z_{\boldsymbol{\rho},t} \sim \mathcal{CN}(0,\sigma^{2})$ and $\mathbf{z}_{j,t} \sim \mathcal{CN}(\mathbf{0},\sigma_{j}^{2}\mathbf{I}_{N_{j}})$ are the zero-mean complex Gaussian noises at the users and the BSs with variance $\sigma^{2}$ and covariance matrix $\sigma_{j}^{2}\mathbf{I}_{N_{j}}$, respectively.

The source symbols of subfile $(f,l)$ for serving request $\boldsymbol{\rho}$ in time slot $t $, denoted by $s_{\boldsymbol{\rho},l,t} \! \in \! \mathbb{C}$, $l \!\in \!\mathcal{L}_{\boldsymbol{\rho}}$, are complex Gaussian random variables with $s_{\boldsymbol{\rho},l,t}\!\sim \! \mathcal{CN}(0, 1)$. Let $\mathbf{w}_{\boldsymbol{\rho},l,t} \!\triangleq\! [\mathbf{w}_{\boldsymbol{\rho},l,0,t}^{H},\ldots \!,\! \mathbf{w}_{\boldsymbol{\rho},l,M,t}^{H}]^{H}$ $\in\mathbb{C}^{N\times1}$ denote the joint beamforming vector for transmit symbol $s_{\boldsymbol{\rho},l,t}$, where $\mathbf{w}_{\boldsymbol{\rho},l,m,t} \in \mathbb{C}^{N_{m}\times1}$ is the individual beamforming vector used by BS $m\in\mathcal{M}$ in time slot $t$. Then, the joint transmit signal of BS set $\mathcal{M}$ in time slot $t \in \mathcal{T}_0$ is given by 
\vspace{-.1cm}
\begin{equation}
\mathbf{x}_{t}=\sum\limits _{\boldsymbol{\rho}\in\mathcal{S}}\,\sum\limits _{l\in\mathcal{L}_{\boldsymbol{\rho}}}\mathbf{w}_{\boldsymbol{\rho},l,t}{s}_{\boldsymbol{\rho},l,t}+\mathbf{v}_{t},
\end{equation}
where superposition coding is used to superimpose the $L_{\boldsymbol{\rho}}$ layers intended for user $\kappa(\boldsymbol{\rho})$ \cite{Tse2005Fundamentals}. Herein, complex Gaussian distributed AN, $\mathbf{v}_{t}\in\mathbb{C}^{N\times1}$, is sent cooperatively by the trusted BSs in set $\mathcal{M}_{\mathcal{T}}$ to proactively interfere the reception of the untrusted BSs in set $\mathcal{M}_{\mathcal{U}}$ \cite{Negi08TWC:AN}. We assume $\mathbf{v}_{t}\sim\mathcal{CN}(\mathbf{0},\,\mathbf{V}_{t})$, where $\mathbf{V}_{t} \in \mathbb{C}^{N \times N} $ is the covariance matrix of the AN, i.e., $\mathbf{V}_{t}\triangleq\mathbb{E}[\mathbf{v}_{t}\mathbf{v}_{t}^{H}] \succeq \mathbf{0}$. As $\mathbf{v}_{t}$ is cooperatively injected  only by the trusted BS set $\mathcal{M_T}$, we require 
$\boldsymbol{\Lambda}_{\mathcal{U}}\mathbf{V}_{t}=\mathbf{0}$, where $\boldsymbol{\Lambda}_{\mathcal{U}}$ is an $N\times N$ diagonal matrix given by $\boldsymbol{\Lambda}_{\mathcal{U}} =\mathrm{diag} \big(\mathbf{0}_{N_{\mathcal{T}}\times1}^{T},\mathbf{1}_{N_{\mathcal{U}}\times1}^{T}\big)$, to ensure that the components of $\mathbf{V}_{t}$ which correspond to untrusted BSs are equal to zero. Moreover, for BS $m\in \mathcal{M}$, participating in cooperative transmission of $s_{\boldsymbol{\rho},l,t}$ is possible only if the requested subfile is cached at the BS, i.e., we require
\vspace{-.2cm}
\begin{alignat}{1}
\textrm{C3}\textrm{:}\;\mathrm{tr}\left(\boldsymbol{\Lambda}_{m} \mathbf{w}_{\boldsymbol{\rho},l,t}\mathbf{w}_{\boldsymbol{\rho},l,t}^{H}\right)\le q_{f(\boldsymbol{\rho}),l,m}P_{m}^{\max}, \qquad\qquad \nonumber \\
 m \in \mathcal{M},\,\boldsymbol{\rho}\in\mathcal{S},\, l\in\mathcal{L}_{\boldsymbol{\rho}}, \, t \in \mathcal{T}_0, \label{eq:cach-bmf}
\end{alignat}
where $P_{m}^{\max}$\! is the maximum transmit power at BS\! $m$, and $\boldsymbol{\Lambda}_{m}$\! is an $N \!\times\! N$ diagonal matrix given by $\boldsymbol{\Lambda}_{m} \!=\! \mathrm{diag}\Big(\mathbf{0}_{(\sum_{j=0}^{m-1}N_{j})\times1}^{T},\mathbf{1}_{N_{m}\times1}^{T},\mathbf{0}_{(\sum_{j=m+1}^{M}N_{j})\times1}^{T}\Big)
$ such that $\mathrm{tr} \big( \boldsymbol{\Lambda}_{m} \mathbf{w}_{\boldsymbol{\rho},l,t} \mathbf{w}_{\boldsymbol{\rho},l,t}^{H} \big) \!\equiv\! \mathrm{tr} \left( \mathbf{w}_{\boldsymbol{\rho},l,m,t} \mathbf{w}_{\boldsymbol{\rho},l,m,t}^{H} \right)$ holds. C3 enforces $\boldsymbol{\Lambda}_{m} \mathbf{w}_{\boldsymbol{\rho},l,t}\mathbf{w}_{\boldsymbol{\rho},l,t}^{H}=\mathbf{0}$, i.e., $\mathbf{w}_{\boldsymbol{\rho},l,m,t} \!=\! \mathbf{0}$, when $q_{f,l,m} \!=\! 0$, i.e., $m \!\notin\! \mathcal{M}_{f,l}^{\mathrm{Coop}}$. Otherwise,  when $q_{f,l,m} \!=\! 1$, i.e., $m \!\in\! \mathcal{M}_{f,l}^{\mathrm{Coop}}$, C3 ensures that the maximum transmit power, $P_{m}^{\max}$, of BS $m \!\in\! \mathcal{M}$ is not exceeded. A constraint of the form of C3 is also referred to as a big-M constraint \cite{Floudas1995MINLP}. Based on C1 and C3, we have $\mathbf{w}_{\boldsymbol{\rho},0,m,t} \!\equiv\! \mathbf{0}$, $\forall m \!\in\! \mathcal{M_{U}}$, i.e., the base-layer subfiles cannot be transmitted by untrusted BSs.

\subsection{Achievable Secrecy Rate}
Each user employs successive interference cancellation (SIC) at the receiver \cite{Tse2005Fundamentals}. The base-layer subfile is decoded first, as it is required for the decoding of the other layers. In decoding the subfile of layer $l\in \mathcal{L}_{\boldsymbol{\rho}} \backslash \left\{ 0\right\} $, the previously decoded lower layers $0,\ldots,l-1$ are first removed from the received signal for interference cancellation. This process continues until layer $L_{\boldsymbol{\rho}}-1$ is decoded \cite{Ohm05:SVC}. Define the interference cancellation coefficient $a_{\boldsymbol{\rho},l}^{\boldsymbol{\rho}',l'}\in\{0,1\}$, where $a_{\boldsymbol{\rho},l}^{\boldsymbol{\rho}',l'}=1$ indicates that the transmission of subfile $\left(f(\boldsymbol{\rho}'),l'\right)$ interferes that of subfile $\left(f(\boldsymbol{\rho}),l\right)$, and $a_{\boldsymbol{\rho},l}^{\boldsymbol{\rho}',l'}=0$ otherwise. By adopting SIC decoding for the SVC video files, we have 
\begin{equation}
a_{\boldsymbol{\rho},l}^{\boldsymbol{\rho}',l'}=
\begin{cases}
0, & \textrm{if }\boldsymbol{\rho}=\boldsymbol{\rho}',\, l\ge l';\\
1, & \textrm{otherwise}.
\end{cases}
\end{equation}
The instantaneous achievable rate (bits/s/Hz) for layer $l\in \mathcal{L}_{\boldsymbol{\rho}}$ at user $\kappa(\boldsymbol{\rho})$ is given by
\begin{alignat}{1}
R_{\boldsymbol{\rho},l,t} & =\log_{2}\left(1+\frac{\frac{1}{\sigma^{2}}\left|\mathbf{h}_{\boldsymbol{\rho},t}^{H}\mathbf{w}_{\boldsymbol{\rho},l,t}\right|^{2}}{1+\frac{1}{\sigma^{2}}I_{\boldsymbol{\rho},l,t}+\frac{1}{\sigma^{2}}\mathbf{h}_{\boldsymbol{\rho},t}^{H}\mathbf{V}_{t}\mathbf{h}_{\boldsymbol{\rho},t}}\right), \label{leg-rate}  \\
I_{\boldsymbol{\rho},l,t} & =\sum\limits_{(\boldsymbol{\rho}',l')\neq(\boldsymbol{\rho},l)}a_{\boldsymbol{\rho},l}^{\boldsymbol{\rho}',l'}\left|\mathbf{h}_{\boldsymbol{\rho},t}^{H}\mathbf{w}_{\boldsymbol{\rho}',l',t}\right|^{2}, 
\end{alignat}
where $I_{\boldsymbol{\rho},l,t}$ is the residual interference term for decoding layer $l$ of user $\kappa(\boldsymbol{\rho})$ and $(\boldsymbol{\rho},l)\neq(\boldsymbol{\rho}',l')$ indicates $\boldsymbol{\rho}\neq\boldsymbol{\rho}'$ and/or $l\neq l'$.

On the other hand, the untrusted BSs may eavesdrop the video information intended for the users. For guaranteeing communication secrecy, the proposed secure delivery scheme is designed to avoid information leakage even under worst-case conditions. Specifically, we assume that BS $j\in\mathcal{M_{U}}$ can fully cancel the self-interference power $\mathbf{c}_{j,t}$ during eavesdropping\footnote{In practice, if self-interference is not perfectly canceled, the residual self-interference impairs the eavesdropping at the untrusted BSs, and hence, improves communication secrecy. However, estimating the residual self-interference at the central controller (e.g., the macro BS), which is responsible for resource allocation, may not be possible. Hence, we make the worst-case assumption of zero self-interference in this paper and the obtained results provide a lower bound on the performance for the case of imperfect self-interference cancellation.}, and hence, achieves the full-duplex capacity upper bound for layer $l$ of the signal intended for user $\kappa(\boldsymbol{\rho})$ given by
\vspace{-.1cm} 
\begin{alignat}{1}
R_{j,\boldsymbol{\rho},l,t} & =\log_{2}\det \left( \mathbf{I}_{N_{j}} + \tfrac{1}{\sigma_{j}^{2}}\mathbf{Z}_{j,\boldsymbol{\rho},l,t}^{-1}\mathbf{G}_{j,t}^{H}\mathbf{w}_{\boldsymbol{\rho},l,t}\mathbf{w}_{\boldsymbol{\rho},l,t}^{H}\mathbf{G}_{j,t} \right), \label{eq:eav-rate} \\
\mathbf{Z}_{j,\boldsymbol{\rho},l,t} & =\mathbf{I}_{N_{j}}+\tfrac{1}{\sigma_{j}^{2}}\mathbf{G}_{j,t}^{H}\mathbf{V}_{t}\mathbf{G}_{j,t}+\tfrac{1}{\sigma_{j}^{2}}\boldsymbol{\Psi}_{j,\boldsymbol{\rho},l,t}\succ\mathbf{0}, \\
\boldsymbol{\Psi}_{j,\boldsymbol{\rho},l,t} & = \!\!\!\!\!\! \sum\limits_{(\boldsymbol{\rho}',l')\neq(\boldsymbol{\rho},l)} \!\!\!\!\!\! a_{\boldsymbol{\rho},l}^{\boldsymbol{\rho}',l'} \left(1-q_{f(\boldsymbol{\rho}'),l',j} \right)\mathbf{G}_{j,t}^{H}\mathbf{w}_{\boldsymbol{\rho}',l',t}\mathbf{w}_{\boldsymbol{\rho}',l',t}^{H}\mathbf{G}_{j,t}. \label{eq:eav-rate2}
\end{alignat}
Note that if subfile $\left(f,l'\right)$ is cached at BS $j\in\mathcal{M_{U}}$, we have $1-q_{f,l',j}=0$ in \eqref{eq:eav-rate2}. That is, in addition to SIC, BS $j\in\mathcal{M_{U}}$ can also utilize the cached video data as side information to suppress the interference caused by subfile $\left(f,l'\right)$. The secrecy rate achievable at user $\kappa(\boldsymbol{\rho})$ for decoding layer $l\in \mathcal{L}_{\boldsymbol{\rho}}$ in time slot $t \in \mathcal{T}_0$ is given by
\vspace{-.1cm} 
\begin{equation}
R_{\boldsymbol{\rho},l,t}^{\mathrm{sec}}=\Big[R_{\boldsymbol{\rho},l,t}-\max_{j\in\mathcal{M_{U}}}R_{j,\boldsymbol{\rho},l,t}\Big]^{+}.
\end{equation}

\begin{rem}
Note that a passive eavesdropper, as considered for non-caching networks in \cite{Negi08TWC:AN,LiTSP11:Opt-Robust,Ng15TWC}  and  caching networks in \cite{Xiang17TWC:SecCaching,LandICC16:HetNets}, can be cast as an untrusted BS having no cache memory or no data cached. Considering C1--C3, such an eavesdropper will not participate in the cooperative transmission of the video files. Thus, the untrusted cache helpers considered in this paper correspond to a more general eavesdropping model than that investigated in the literature \cite{Xiang17TWC:SecCaching,LandICC16:HetNets,Negi08TWC:AN,LiTSP11:Opt-Robust,Ng15TWC}. 
\end{rem}

\section{\label{sec:Problem-Formul}Problem Formulation }
In this section, we first present the adopted imperfect CSI model for video delivery. Then, a two-timescale robust optimization problem is formulated for minimization of the total BS transmit power required for video streaming under QoS and secrecy constraints. Note that low transmit power is desirable to minimize the interference caused in other cells and to reduce the network operation cost. For a given cache status, the cooperative transmission decisions for each time slot are optimized online based on instantaneous CSI estimates. However, due to time causality and computational complexity constraints, the cache placement for each period is optimized offline based on historical user requests and CSI \cite{Xiang17TVT:CLCaching,Xiang17TWC:SecCaching}.

\subsection{Channel State Information}
At the beginning of each time slot, the CSI $\mathbf{h}_{\boldsymbol{\rho},t}$ and $\mathbf{G}_{j,t}$ has to be acquired\footnote{For example, by exploiting channel reciprocity in time division duplex systems, $\mathbf{h}_{\boldsymbol{\rho},t}$ and $\mathbf{G}_{j,t}$ can be estimated in the uplink at the small cell and macro BSs based on pilots emitted by the UEs and the untrusted BSs, respectively. Then, the estimated CSI obtained at the small cell BSs is fed back to the macro BS via the X2 interface \cite{IEEE3GPP:TR36814}.} at the centralized controller, i.e., the macro BS, for computing the resource allocation. The estimates of $\mathbf{h}_{\boldsymbol{\rho},t}$ and $\mathbf{G}_{j,t}$ gathered at the macro BS, denoted by $\widehat{\mathbf{h}}_{\boldsymbol{\rho},t} \in \mathbb{C}^{N\times1}$ and $\widehat{\mathbf{G}}_{j,t} \in \mathbb{C}^{N\times N_{j}}$, respectively, will in general be imperfect. That is, the actual channels are given by $\mathbf{h}_{\boldsymbol{\rho},t} =\widehat{\mathbf{h}}_{\boldsymbol{\rho},t} + \Delta\mathbf{h}_{\boldsymbol{\rho},t}$ and $\mathbf{G}_{j,t} = \widehat{\mathbf{G}}_{j,t}+\Delta\mathbf{G}_{j,t}$, where $\Delta\mathbf{h}_{\boldsymbol{\rho},t}$ and $\Delta\mathbf{G}_{j,t}$ represent the respective channel estimation errors caused by quantization errors, imperfect feedback channels, as well as outdated and noisy estimates. In fact, the estimation errors $\Delta\mathbf{h}_{\boldsymbol{\rho},t}$ and $\Delta\mathbf{G}_{j,t}$ may be enhanced by the actions of the untrusted BSs which may not fully cooperate with the macro BS during channel estimation and feedback.

The specific values of $\Delta\mathbf{h}_{\boldsymbol{\rho},t}$ and $\Delta\mathbf{G}_{j,t}$ are not known at the macro BS. To model the imperfect CSI, we assume that the possible values of $\Delta\mathbf{h}_{\boldsymbol{\rho},t}$ and $\Delta\mathbf{G}_{j,t}$ lie in ellipsoidal uncertainty regions \cite{palomar2010convex} given by
\begin{alignat}{1}
\boldsymbol{\Omega}_{\boldsymbol{\rho},t} & \triangleq\Big\{ \Delta\mathbf{h}_{\boldsymbol{\rho},t}\in\mathbb{C}^{N\times1}\mid\Delta\mathbf{h}_{\boldsymbol{\rho},t}^{H}\boldsymbol{\Xi}_{\boldsymbol{\rho}}\Delta\mathbf{h}_{\boldsymbol{\rho},t}\le\varepsilon_{\boldsymbol{\rho}}^{2}\Big\} , \nonumber \\ & \qquad\qquad\qquad\qquad\qquad\qquad \boldsymbol{\rho}\in\mathcal{S},\, t \in \mathcal{T}_0, \\
\mathcal{\boldsymbol{\Omega}}_{j,t} & \triangleq\Big\{ \Delta\mathbf{G}_{j,t}\in\mathbb{C}^{N\times N_{j}}\mid\mathrm{tr}\left(\Delta\mathbf{G}_{j,t}^{H}\boldsymbol{\boldsymbol{\Xi}}_{j}\Delta\mathbf{G}_{j,t}\right)\le\varepsilon_{j}^{2}\Big\} , \nonumber \\
& \qquad\qquad\qquad\qquad\qquad\qquad j\in\mathcal{M_{U}},\, t \in \mathcal{T}_0. 
\end{alignat}
Here, $\varepsilon_{\boldsymbol{\rho}}>0$ and $\varepsilon_{j}>0$ represent the radii of uncertainty regions $\boldsymbol{\Omega}_{\boldsymbol{\rho},t}$ and $\mathcal{\boldsymbol{\Omega}}_{j,t}$, respectively; $\boldsymbol{\Xi}_{\boldsymbol{\rho}} \in \mathbb{C}^{N\times N}$ and $\boldsymbol{\boldsymbol{\Xi}}_{j} \in \mathbb{C}^{N\times N}$ denote the orientations of the uncertainty regions, respectively, where $\boldsymbol{\Xi}_{\boldsymbol{\rho}} \succ \mathbf{0}$ and $\boldsymbol{\boldsymbol{\Xi}}_{j} \succ \mathbf{0}$. In practice, the values of $\varepsilon_{\boldsymbol{\rho}}$, $\varepsilon_{j}$, $\boldsymbol{\Xi}_{\boldsymbol{\rho}}$, and $\boldsymbol{\boldsymbol{\Xi}}_{j}$ depend on the channel coherence time and the adopted channel estimation methods.

\subsection{Caching Optimization}
Let $\mathbf{q}\triangleq \left[q_{1,0,0},\ldots,q_{f,l,m},\ldots,q_{F,L-1,M}\right]$ and $\mathbf{w}_{\boldsymbol{\rho},t} \triangleq [ \mathbf{w}_{\boldsymbol{\rho},0,t}^H, \ldots, \mathbf{w}_{\boldsymbol{\rho},L_{\boldsymbol{\rho}}-1,t}^H ]^H$ be the caching and the transmitter beamforming optimization vectors, respectively. Considering the two-timescale control in Fig.~\ref{fig1b}, the caching decision $\mathbf{q}$ is made (at the end of) every $T_0$ time slots based on the historical profiles of user requests and CSI that have been collected during the time period\footnote{Prediction of the users' future requests based on historical user profiles can further improve the cache placement at the cost of an increased computational complexity.}. For the considered typical period $\mathcal{T}_0$, the caching optimization problem is formulated as:
\begin{align}
\textrm{P0:}\;\minimize_{\mathbf{q},\mathbf{w}_{\boldsymbol{\rho},t},\mathbf{V}_{t}}\; & \sum\limits_{t\in\mathcal{T}_{0}}U_{\textrm{TP}}\left(\mathbf{q},\mathbf{w}_{\boldsymbol{\rho},t},\mathbf{V}_{t}\right)\label{eq:R1}\\
\st \;\;& \textrm{C1, C2, C3, }\textrm{C4: }\mathbf{V}_{t}\succeq\mathbf{0},\,\boldsymbol{\Lambda}_{\mathcal{U}}\mathbf{V}_{t}=\mathbf{0}, \nonumber \\ 
 & \textrm{C5: }\mathrm{tr} \left(\! \boldsymbol{\Lambda}_{m} \left(\! \sum_{\boldsymbol{\rho},l}\mathbf{w}_{\boldsymbol{\rho},l,t}\mathbf{w}_{\boldsymbol{\rho},l,t}^{H} \!+\! \mathbf{V}_{t} \right) \right) \!\le\! P_{m}^{\max}, \nonumber \\
&\textrm{C6:}\min_{\Delta\mathbf{h}_{\boldsymbol{\rho},t}\in\boldsymbol{\Omega}_{\boldsymbol{\rho},t}}R_{\boldsymbol{\rho},l,t}\ge R_{\boldsymbol{\rho},l}^{\textrm{req}},\,\boldsymbol{\rho}\in\mathcal{S},\; l\in \mathcal{L}_{\boldsymbol{\rho}},\nonumber \\ 
&\textrm{C7:}\max_{j\in\mathcal{M_{U}}}\max_{\Delta\mathbf{G}_{j,t}\in\mathcal{\boldsymbol{\Omega}}_{j,t}}R_{j,\boldsymbol{\rho},0,t}\le R_{\boldsymbol{\rho},0}^{\mathrm{tol}},\,\boldsymbol{\rho}\in\mathcal{S}, \nonumber 
\end{align}
where $U_{\textrm{TP}}\left(\mathbf{q},\mathbf{w}_{\boldsymbol{\rho},t},\mathbf{V}_{t}\right) \triangleq \mathrm{tr} \left( \sum_{m \in \mathcal{M}} \boldsymbol{\Lambda}_{m} \left(\sum_{\boldsymbol{\rho}\in\mathcal{S}}\sum_{l\in \mathcal{L}_{\boldsymbol{\rho}}} \mathbf{w}_{\boldsymbol{\rho},l,t} \mathbf{w}_{\boldsymbol{\rho},l,t}^{H} + \mathbf{V}_{t} \right) \right)$ denotes the total BS transmit power in time slot $t \in \mathcal{T}_{0}$. Constraint C5 limits the maximal transmit power of BS $m\in\mathcal{M}$ to $P_{m}^{\max}$. C6 guarantees the minimum video delivery rate, $R_{\boldsymbol{\rho},l}^{\textrm{req}}$, in each time slot $t\in\mathcal{T}_{0}$ to provide  QoS in delivering layer $l\in\mathcal{L}_{\boldsymbol{\rho}}$ for serving request $\boldsymbol{\rho}\in\mathcal{S}$. C7 constrains the maximum data rate leaked to the untrusted BSs in set $\mathcal{M_{U}}$ to $R_{\boldsymbol{\rho},0}^{\mathrm{tol}}$ in each time slot $t$ to ensure communication secrecy. Since the untrusted BSs are unable to decode the enhancement layers without base-layer information, secrecy can be ensured by imposing C7 only on the delivery of the base-layer subfiles. Due to the imperfect CSI, the minimum/maximum data rate in C6/C7 is guaranteed/constrained for all possible estimation error vectors/matrices in the respective uncertainty sets in order to facilitate robustness with respect to (w.r.t.) communication secrecy. This robust optimization approach has been commonly adopted for studying PLS in the literature, see \cite{Negi08TWC:AN,LiTSP11:Opt-Robust,Ng15TWC} and references therein. Constraints C6 and C7 jointly guarantee a minimum achievable secrecy rate of $R_{\boldsymbol{\rho},0,t}^{\mathrm{sec}} = \left[R_{\boldsymbol{\rho},0}^{\textrm{req}}-R_{\boldsymbol{\rho},0}^{\mathrm{tol}}\right]^{+}$, $t\in\mathcal{T}_{0}$, for delivering the base-layer subfiles for request $\boldsymbol{\rho}$, provided that problem P0 is feasible.

Problem P0 is a non-convex mixed-integer nonlinear program (MINLP)%
\footnote{For a non-convex MINLP, even if the binary constraints are relaxed into convex ones, the problem remains non-convex \cite{Floudas1995MINLP}.%
} due to the binary caching decision vector $\mathbf{q}$ and the non-convex constraints C6 and C7. This type of problem is NP-hard \cite{Floudas1995MINLP}. Yet, since P0 is solved offline for a large timescale, we adopt a global optimization method to solve P0 optimally in Section~\ref{sec-iv-a}.  The obtained solution defines a performance benchmark for low-complexity suboptimal schemes, cf. Section~\ref{sec-iv-b}.

\subsection{Delivery Optimization}
Assume that the instantaneous CSI estimates are given. Moreover, the cache status $\mathbf{q}$ for $\mathcal{T}_{0}$ has been determined at the end of the previous time period. Then, the cooperative transmission policy $\left\{ \mathbf{w}_{\boldsymbol{\rho},t},\mathbf{V}_{t}\right\} $ for time $t\in\mathcal{T}_{0}$ is optimized online by solving the following problem
\vspace{-.1cm}
\begin{align}
\textrm{Q0:}\;\minimize_{\mathbf{w}_{\boldsymbol{\rho},t},\mathbf{V}_{t}}\quad & U_{\textrm{TP}}\left(\mathbf{q},\mathbf{w}_{\boldsymbol{\rho},t},\mathbf{V}_{t}\right)\label{eq:R2}\\
\st\quad & \textrm{C3, }\textrm{C4, C5, C6, C7.}\nonumber 
\end{align}
Problem Q0 is non-convex due to constraints C6 and C7. However, we will show that Q0 can be optimally solved by employing SDP relaxation, cf. Section~\ref{sec-iv-c}.

\section{\label{sec:Problem-Sol}Problem Solution}
In this section, the caching problem~P0 is tackled first. We show that P0 can be transformed into a convex MINLP by SDP relaxation and further solved optimally by an iterative algorithm. Inspired by the optimal algorithm, a low-complexity suboptimal caching scheme is developed to balance between optimality and computational complexity. Moreover, we show that the delivery problem~Q0 can be optimally and efficiently solved. 

\subsection{\label{sec-iv-a}Optimal Caching Scheme} 
\subsubsection{Problem Transformation}
To reformulate problem~P0 as a convex MINLP, C6 and C7 have to be transformed into convex constraints. Let $\mathbf{W}_{\boldsymbol{\rho},l,t} = \mathbf{w}_{\boldsymbol{\rho},l,t} \mathbf{w}_{\boldsymbol{\rho},l,t}^{H} \succeq \mathbf{0}$ be the beamforming matrix subject to $\mathrm{rank} \left(\mathbf{W}_{\boldsymbol{\rho},l,t}\right) \le 1$.  By substituting $\mathbf{W}_{\boldsymbol{\rho},l,t}$ and employing elementary arithmetic operations, C6 is equivalently reformulated as an affine inequality constraint that is jointly convex w.r.t. $\left\{ \mathbf{W}_{\boldsymbol{\rho},l,t},\mathbf{V}_{t}\right\} $,
\vspace{-.1cm}
\begin{equation}
\overline{\textrm{C6}}\textrm{: }\mathbf{h}_{\boldsymbol{\rho},t}^{H}\mathbf{T}_{\boldsymbol{\rho},l,t}\mathbf{h}_{\boldsymbol{\rho},t}\ge\sigma^{2},\quad\forall\Delta\mathbf{h}_{\boldsymbol{\rho},t}\in\boldsymbol{\Omega}_{\boldsymbol{\rho},t},
\end{equation}
where $\mathbf{T}_{\boldsymbol{\rho},l,t}\triangleq\tfrac{1}{\eta_{\boldsymbol{\rho},l}^{\textrm{req}}} \mathbf{W}_{\boldsymbol{\rho},l,t} - \sum_{(\boldsymbol{\rho}',l') \neq (\boldsymbol{\rho},l)} a_{\boldsymbol{\rho}, l}^{\boldsymbol{\rho}',l'} \mathbf{W}_{\boldsymbol{\rho}',l',t} - \mathbf{V}_{t}$
and $\eta_{\boldsymbol{\rho},l}^{\textrm{req}} \triangleq 2^{R_{\boldsymbol{\rho},l}^{\textrm{req}}}-1$. However, as $\boldsymbol{\Omega}_{\boldsymbol{\rho},t}$ is a continuous set, $\overline{\textrm{C6}}$ is semi-infinite, i.e., it represents infinitely many inequalities for $\mathbf{T}_{\boldsymbol{\rho},l,t}$, and hence, is still intractable for optimization. To overcome this issue, $\overline{\textrm{C6}}$ is transformed into a finite number of convex constraints. To this end, we substitute $\mathbf{h}_{\boldsymbol{\rho},t} = \widehat{\mathbf{h}}_{\boldsymbol{\rho},t} +\Delta\mathbf{h}_{\boldsymbol{\rho},t}$ in $\overline{\textrm{C6}}$ and apply the S-procedure from \cite[Appendix B]{Boyd2004Convex}, which leads to
\begin{alignat}{1}
 \overline{\textrm{C6}}\textrm{: } & \Delta\mathbf{h}_{\boldsymbol{\rho},t}^{H} \mathbf{T}_{\boldsymbol{\rho},l,t}\Delta \mathbf{h}_{\boldsymbol{\rho},t} + 2\Re\Big\{ \widehat{\mathbf{h}}_{\boldsymbol{\rho},t}^{H}\mathbf{T}_{\boldsymbol{\rho},l,t} \Delta \mathbf{h}_{\boldsymbol{\rho},t}\Big\} \nonumber \\
& \quad + \widehat{\mathbf{h}}_{\boldsymbol{\rho},t}^{H}\mathbf{T}_{\boldsymbol{\rho},l,t} \widehat{\mathbf{h}}_{\boldsymbol{\rho},t}-\sigma^{2}\ge0,\quad\forall\Delta\mathbf{h}_{\boldsymbol{\rho},t}\in\boldsymbol{\Omega}_{\boldsymbol{\rho},t},\nonumber \\
\iff \widetilde{\textrm{C6}} & \textrm{: }\mathbf{U}_{\boldsymbol{\rho},t}^{H}\mathbf{T}_{\boldsymbol{\rho},l,t}\mathbf{U}_{\boldsymbol{\rho},t} \succeq \nonumber \\
&  \quad \Bigg[\begin{array}{cc}
-\delta_{\boldsymbol{\rho},l,t}\boldsymbol{\Xi}_{\boldsymbol{\rho}} & \mathbf{0}\\
\mathbf{0}^{H} & \sigma^{2}+\delta_{\boldsymbol{\rho},l,t}\varepsilon_{\boldsymbol{\rho}}^{2}
\end{array}\Bigg],\; \delta_{\boldsymbol{\rho},l,t}\ge0,
\end{alignat}
where $\mathbf{U}_{\boldsymbol{\rho},t} \triangleq [\mathbf{I}_{N},\widehat{\mathbf{h}}_{\boldsymbol{\rho},t}] \in \mathbb{C}^{(N+1)\times(N+1)}$ and $\delta_{\boldsymbol{\rho},l,t}$ is an auxiliary optimization variable.

Next, let $\overline{\mathbf{W}}_{\boldsymbol{\rho},l,j,t} = (1-q_{f(\boldsymbol{\rho}),l,j}) \mathbf{W}_{\boldsymbol{\rho},l,t} \succeq \mathbf{0}$ be an auxiliary optimization matrix. We have 
\begin{equation}
\boldsymbol{\Psi}_{j,\boldsymbol{\rho},l,t}=\mathbf{G}_{j,t}^{H}\Big(\sum\limits _{(\boldsymbol{\rho}',l')\neq(\boldsymbol{\rho},l)}a_{\boldsymbol{\rho},l}^{\boldsymbol{\rho}',l'}\overline{\mathbf{W}}_{\boldsymbol{\rho}',l',j,t}\Big)\mathbf{G}_{j,t},
\end{equation}
if and only if $\mathrm{rank}(\overline{\mathbf{W}}_{\boldsymbol{\rho},l,j,t})\le1$ and the following constraints hold, 
\begin{equation}
\begin{alignedat}{1}\textrm{C8:}\; & \mathrm{tr}(\mathbf{W}_{\boldsymbol{\rho},l,t}-\overline{\mathbf{W}}_{\boldsymbol{\rho},l,j,t})\preceq q_{f(\boldsymbol{\rho}),l,j}P_{\max},  \\
\textrm{C9:}\; & \mathrm{tr}(\overline{\mathbf{W}}_{\boldsymbol{\rho},l,j,t})\preceq(1-q_{f(\boldsymbol{\rho}),l,j})P_{\max},\\
\textrm{C10:}\; & \mathbf{W}_{\boldsymbol{\rho},l,t}\succeq\overline{\mathbf{W}}_{\boldsymbol{\rho},l,j,t},\;\overline{\mathbf{W}}_{\boldsymbol{\rho},l,j,t}\succeq\mathbf{0}, 
\end{alignedat}
\end{equation}
where $P_{\max}\triangleq\sum_{m\in\mathcal{M}}P_{m}^{\max}$. Here, C8 and C9 guarantee that $\overline{\mathbf{W}}_{\boldsymbol{\rho},l,j,t}=\mathbf{0}$ if $q_{f(\boldsymbol{\rho}),l,j}=1$, and $\overline{\mathbf{W}}_{\boldsymbol{\rho},l,j,t}=\mathbf{W}_{\boldsymbol{\rho},l,t}$ otherwise. By substituting $\overline{\mathbf{W}}_{\boldsymbol{\rho},l,j,t}$ and $\boldsymbol{\Psi}_{j,\boldsymbol{\rho},l,t}$, C7 can be reformulated into an LMI as follows 
\vspace{-.2cm}
\begin{alignat}{1}
\textrm{C7} & \iff\tfrac{1}{\sigma_{j}^{2}}\mathbf{w}_{\boldsymbol{\rho},0,t}^{H}\mathbf{G}_{j,t}\mathbf{Z}_{j,\boldsymbol{\rho},0,t}^{-1}\mathbf{G}_{j,t}^{H}\mathbf{w}_{\boldsymbol{\rho},0,t}\le \eta_{\boldsymbol{\rho},0}^{\textrm{tot}} \nonumber \\
  &\iff\mathrm{tr} \left(\mathbf{Z}_{j,\boldsymbol{\rho},0,t}^{-1}\mathbf{G}_{j,t}^{H}\mathbf{W}_{\boldsymbol{\rho},0,t}\mathbf{G}_{j,t} \right)\le\sigma_{j}^{2}\eta_{\boldsymbol{\rho},0}^{\textrm{tot}},\nonumber \\
 & \overset{\textrm{(a)}}{\iff}\lambda_{\max} \left(\mathbf{Z}_{j,\boldsymbol{\rho},0,t}^{-\nicefrac{1}{2}}\mathbf{G}_{j,t}^{H}\mathbf{W}_{\boldsymbol{\rho},0,t}\mathbf{G}_{j,t}\mathbf{Z}_{j,\boldsymbol{\rho},0}^{-\nicefrac{1}{2}} \right)\le\sigma_{j}^{2}\eta_{\boldsymbol{\rho},0}^{\textrm{tot}},\nonumber \\
 & \iff\overline{\textrm{C7}}\textrm{: }\mathbf{G}_{j,t}^{H}\mathbf{T}_{\boldsymbol{\rho},0,j,t}\mathbf{G}_{j,t}\preceq\sigma_{j}^{2}\mathbf{I}_{N_{j}},\;\forall j\in\mathcal{M_{U}}, \nonumber \\
& \qquad\qquad\qquad\qquad\qquad\qquad  \forall\Delta\mathbf{G}_{j,t}\in\mathcal{\boldsymbol{\Omega}}_{j,t},\label{eq:C7}
\end{alignat}
where $\eta_{\boldsymbol{\rho},0}^{\textrm{tot}} \triangleq 2^{R_{\boldsymbol{\rho},0}^{\textrm{tot}}}-1$,  $\mathbf{T}_{\boldsymbol{\rho},0,j,t}\triangleq\frac{1}{\eta_{\boldsymbol{\rho},0}^{\textrm{tot}}} \mathbf{W}_{\boldsymbol{\rho},0,t} - \sum\nolimits _{(\boldsymbol{\rho}',l') \neq (\boldsymbol{\rho},0)} a_{\boldsymbol{\rho},0}^{\boldsymbol{\rho}',l'} \overline{\mathbf{W}}_{\boldsymbol{\rho}',l',j,t} - \mathbf{V}_{t}$, and (a) holds due to $\mathrm{rank}(\overline{\mathbf{W}}_{\boldsymbol{\rho},l,t}) \le 1$. In fact, as $\boldsymbol{\Omega}_{j,t}$ is a continuous set, $\overline{\textrm{C7}}$ in \eqref{eq:C7} represents infinitely many LMIs that are jointly convex w.r.t. $\left\{ \mathbf{W}_{ \boldsymbol{\rho},0,t}, \mathbf{V}_{t}, \overline{\mathbf{W}}_{j,t} \right\}$. For tractability, $\overline{\textrm{C7}}$ has to be transformed into a finite number of convex constraints. This can be accomplished by exploiting the robust quadratic matrix inequality in \cite[Theorem 3.3]{Luo04quadratic}. Thereby, we obtain
\begin{alignat}{1}
\overline{\textrm{C7}}  & \iff \widetilde{\textrm{C7}}\textrm{: }\mathbf{U}_{j,t}^{H}\mathbf{T}_{\boldsymbol{\rho},0,j,t}\mathbf{U}_{j,t}\preceq \nonumber \\ 
& \left[\begin{array}{cc}
(\sigma_{j}^{2}-\delta_{\boldsymbol{\rho},0,j,t})\mathbf{I}_{N_{j}} & \mathbf{0}\\
\mathbf{0} & \frac{\delta_{\boldsymbol{\rho},0,j,t}}{\varepsilon_{j}^{2}}\boldsymbol{\Xi}_{j}
\end{array}\right],\delta_{\boldsymbol{\rho},0,j,t}\ge0,
\label{newC7}
\end{alignat}
where $\mathbf{U}_{j,t} \triangleq [\widehat{\mathbf{G}}_{j,t} \mathbf{I}_{N}] \in \mathbb{C}^{N\times(N+N_{j})}$ and $\delta_{\boldsymbol{\rho},0,j,t}$ is an auxiliary optimization variable.

Finally, by defining the delivery variable $\mathbf{D}_{t} \triangleq [\mathbf{W}_{\boldsymbol{\rho},l,t},\overline{\mathbf{W}}_{\boldsymbol{\rho},l,j,t},\mathbf{V}_{t}]$ and applying the above transformations, the original problem P0 is equivalently reformulated as 
\begin{align}
 \minimize_{\mathbf{q},\mathbf{D}_{t}}\quad & \sum\limits _{t\in\mathcal{T}_{0}}U_{\textrm{TP}}\left(\mathbf{q},\mathbf{D}_{t}\right)\label{eq:R1-2}\\
\st \quad & \textrm{C1, C2, C4, \ensuremath{\widetilde{\textrm{C6}}}, \ensuremath{\widetilde{\textrm{C7}}}, C8, C9, C10,} \nonumber \\
 & \textrm{C3}\textrm{: }\mathrm{tr}(\boldsymbol{\Lambda}_{m}\mathbf{W}_{\boldsymbol{\rho},l,t})\le q_{f(\boldsymbol{\rho}),l,m}P_{m}^{\max}, \nonumber \\
& \textrm{C5: }\mathrm{tr}\bigg(\boldsymbol{\Lambda}_{m} \Big(\sum\limits _{\boldsymbol{\rho},l}\mathbf{W}_{\boldsymbol{\rho},l,t}+\mathbf{V}_{t} \Big) \bigg)\le P_{m}^{\max},\nonumber \\
 & \textrm{\textrm{C11}\textrm{:} }\mathrm{rank}(\mathbf{W}_{\boldsymbol{\rho},l,t})\le1,\;\boldsymbol{\rho}\in\mathcal{S},\, l\in\mathcal{L}_{\boldsymbol{\rho}}.\quad\nonumber 
\end{align}
Here, constraint $\mathrm{rank} (\overline{\mathbf{W}}_{\boldsymbol{\rho},l,t}) \le1$ is dropped due to C10 and C11. Let P1 denote the SDP relaxation of problem \eqref{eq:R1-2}, obtained by dropping C11 in \eqref{eq:R1-2}. Then, problem P1 is a convex MINLP, i.e., by relaxing the binary constraints of P1 into convex ones, we arrive at a convex problem.

The GBD algorithm is a simple iterative method to handle convex MINLPs \cite[Section 6.3]{Floudas1995MINLP}. In each GBD iteration, upper and lower bounds on the optimal value are generated by solving a primal subproblem and a master problem, respectively. To ensure convergence, optimality and feasibility cuts are successively added to tighten the bounds and eliminate the infeasible solutions possibly obtained during the iterations, respectively. The GBD algorithm is attractive for solving P1 as it can be efficiently implemented exploiting the structure of P1. In particular, the resulting primal subproblem is a convex problem where strong duality holds while the master problem is a mixed-integer linear program (MILP), and both problems are easy to handle using off-the-shelf numerical solvers such as CVX \cite{CVX} and MOSEK \cite{MOSEK}. However, the GBD algorithm typically suffers from slow convergence. This is because, when an infeasible solution is obtained in an iteration of the GBD, the resulting feasibility cut is usually ineffective in improving the solution. If the problem is infeasible, the GBD algorithm terminates only after having performed an exhaustive search over all possible candidate solutions. To remedy this issue, an improved GBD algorithm\footnote{A similar approach as in the proposed improved GBD algorithm has been successfully applied to accelerate the outer approximation algorithm in \cite[Section 6.6]{Floudas1995MINLP}.} is proposed below.

\subsubsection{Problem Decomposition} 
The proposed modified GBD algorithm applies a two-layer decomposition of problem P1 and solves a binary caching optimization problem for $\mathbf{q}$ in the outer layer and a continuous delivery optimization problem for $\mathbf{D}_{t}$ in the inner layer. However, $\mathbf{q}$ and $\mathbf{D}_{t}$ are coupled via constraints C3, C8, and C9. To facilitate the decomposition, we perturb the right-hand sides of C3, C8, and C9 by introducing slack variables $s_{\boldsymbol{\rho},l,m,t}^{\textrm{C3}} \ge 0$, $s_{\boldsymbol{\rho},l,j,t}^{\textrm{C8}} \ge0$, and $s_{\boldsymbol{\rho},l,j,t}^{\textrm{C9}} \ge 0$, respectively. Let $\mathbf{s}_{t} \triangleq [s_{\boldsymbol{\rho},l,m,t}^{\textrm{C3}},\, s_{\boldsymbol{\rho},l,j,t}^{ \textrm{C8}}, \, s_{\boldsymbol{\rho},l,j,t}^{\textrm{C9}}]$ be the perturbation vector and $\mathbf{s}_{t}\succeq \mathbf{0}$. Moreover, in the objective function, we add an $\ell_{1}$-norm (exact) penalty cost function for $\mathbf{s}_{t}$, 
\vspace{-.2cm}
\begin{equation}
f_{\textrm{Pen}}(\mathbf{s}_{t}) \! \triangleq\! \mu\left(\sum\limits _{\boldsymbol{\rho},l,m} s_{\boldsymbol{\rho},l,m,t}^{\textrm{C3}} + \sum\limits _{\boldsymbol{\rho},l,j} \left(s_{\boldsymbol{\rho},l,j,t}^{\textrm{C8}} + s_{\boldsymbol{\rho},l,j,t}^{\textrm{C9}} \right)\right),
\end{equation}
with penalty factor $\mu\gg1$. Consequently, problem P1 decomposes into $T_{0}$ SDP subproblems in the inner layer, i.e., one subproblem for each time slot $t\in\mathcal{T}_{0}$, and an MILP in the outer layer, which are shown in \eqref{eq:Primal} and \eqref{eq:Master} at the top of the next page, respectively.
\begin{figure*}
\begin{align}
\nu_{t}\left(\mathbf{q}\right)\triangleq\minimize_{\mathbf{D}_{t},\mathbf{s}_{t}\succeq \mathbf{0}}\; & U_{\textrm{TP}}(\mathbf{q},\mathbf{D}_{t})+f_{\textrm{Pen}}(\mathbf{s}_{t}) \label{eq:Primal}\\
\st \; & \mathbf{D}_{t} \! \in \! \mathcal{D}_{t} \!\triangleq\! \left\{ \mathbf{D}_{t} \!\mid\! \textrm{C4,\! C5,\! \ensuremath{\overline{\textrm{C6}}},\! \ensuremath{\overline{\textrm{C7}}},\! C10}\right\}\!, 
\qquad \overline{\textrm{C3}}\textrm{:}\, \mathrm{tr}(\boldsymbol{\Lambda}_{m}\mathbf{W}_{\boldsymbol{\rho},l,t}) \!-\! q_{f(\boldsymbol{\rho}),l,m}P_{m}^{\max} \!\le\! s_{\boldsymbol{\rho},l,m}^{\textrm{C3}},\nonumber \\
 & \overline{\textrm{C8}}\textrm{: }\mathrm{tr}(\mathbf{W}_{\boldsymbol{\rho},l,t}-\overline{\mathbf{W}}_{\boldsymbol{\rho},l,j,t})-q_{f(\boldsymbol{\rho}),l,j}P_{\max}\le s_{\boldsymbol{\rho},l,j}^{\textrm{C8}}, 
\qquad  \overline{\textrm{C9}}\textrm{: }\mathrm{tr}(\overline{\mathbf{W}}_{\boldsymbol{\rho},l,j,t})-(1-q_{f(\boldsymbol{\rho}),l,j})P_{\max}\le s_{\boldsymbol{\rho},l,j}^{\textrm{C9}},\nonumber 
\\
\minimize_{\mathbf{q},\alpha}\quad & \alpha\label{eq:Master}\\
\st \quad & \alpha\ge\nu\left(\mathbf{q}\right) \triangleq \sum\nolimits _{t \in \mathcal{T}_{0}} \nu_{t} \left(\mathbf{q}\right), \qquad \mathbf{q} \in \mathcal{Q} \triangleq \left\{ \mathbf{q} \mid \textrm{C1, C2}\right\}.\nonumber 
\end{align}
\hrulefill
\end{figure*}
Problem \eqref{eq:Master} is referred as the master problem. Thereby, problems \eqref{eq:Primal} and \eqref{eq:Master}  are equivalent to P1 when $\mu \gg 1$, as stated in Proposition~\ref{prop1}. 
\begin{prop}
\emph{\label{prop1}For $\mu\gg1$, problems P1 and \eqref{eq:Primal}, \eqref{eq:Master} are equivalent such that: i) if P1 is feasible, then SDP subproblem \eqref{eq:Primal} is \emph{always} feasible for $\mathbf{q}\in\mathcal{Q}$; moreover, the optimal solution of $\mathbf{q}$ for P1 solves the master problem \eqref{eq:Master}; ii) if  problem \eqref{eq:Primal} is infeasible, i.e., $\nu\left(\mathbf{q}\right)=+\infty$, or its optimal solution satisfies $\mathbf{s}_{t'}\neq\mathbf{0}$ for some $t'\in\mathcal{T}_{0}$, then problem P1 is infeasible.} 
\end{prop}
\begin{IEEEproof}
Please refer to Appendix~\ref{proof1}. 
\end{IEEEproof}
\begin{rem} \label{rem3}
By perturbation, the feasible set of $\nu \left( \mathbf{q} \right)$ in \eqref{eq:Primal}, \eqref{eq:Master} is extended to $\mathcal{Q} \subseteq \left\{ 0,1\right\} ^{F \times L\times M}$. Consequently, based on Proposition~\ref{prop1}, infeasible solutions can be avoided if problem P1 is feasible, cf. i), or identified easily if problem P1 is infeasible, cf. ii). These properties facilitate an efficient implementation of the GBD algorithm in the sequel to optimally solve \eqref{eq:Primal} and \eqref{eq:Master}. 
\end{rem}
  SDP subproblem \eqref{eq:Primal} is convex and can be solved by interior point methods \cite{Ye2011interior,Boyd2004Convex}, i.e., numerical solvers such as CVX \cite{CVX} are applicable. Meanwhile, exploiting the convexity (and strong duality) of  \eqref{eq:Primal}, we can further simplify the formulation of the master problem \eqref{eq:Master}. Let $\lambda_{\boldsymbol{\rho},l,m,t}^{\overline{\textrm{C3}}} \ge 0 $, $\lambda_{\boldsymbol{\rho},l,j,t}^{\overline{\textrm{C8}}} \ge 0$, and $\lambda_{\boldsymbol{\rho},l,j,t}^{\overline{\textrm{C9}}} \ge 0$ be the Lagrange multipliers for $\overline{\textrm{C3}}$, $\overline{\textrm{C8}}$, and $\overline{\textrm{C9}}$, respectively, and define $\boldsymbol{\lambda}_{t} \!\triangleq\! [\lambda_{\boldsymbol{\rho},l,m,t}^{\overline{\textrm{C3}}}, \lambda_{\boldsymbol{\rho},l,j,t}^{\overline{\textrm{C8}}}, \lambda_{\boldsymbol{\rho},l,j,t}^{\overline{\textrm{C9}}}] \!\succeq\! \mathbf{0}$. The Lagrangian of~\eqref{eq:Primal} can be written as 
\vspace{-.2cm}
\begin{equation}
\begin{alignedat}{1}\mathcal{L}_{\mathbf{q}}\left(\mathbf{D}_{t},\mathbf{s}_{t};\boldsymbol{\lambda}_{t}\right) & =f_{1}\left(\mathbf{q};\boldsymbol{\lambda}_{t}\right)+f_{2}\left(\mathbf{D}_{t},\mathbf{s}_{t};\boldsymbol{\lambda}_{t}\right),\end{alignedat}
\label{eq:Lagrang}
\end{equation}
where $f_{1}\left(\mathbf{q};\boldsymbol{\lambda}_{t}\right) = \sum_{\boldsymbol{\rho},l,j} \big( \lambda_{\boldsymbol{\rho},l,j,t}^{\overline{\textrm{C9}}} - \lambda_{\boldsymbol{\rho},l,j,t}^{\overline{\textrm{C8}}} \big) q_{f(\boldsymbol{\rho}),l,j} P_{\max} - \sum_{\boldsymbol{\rho},l,m,t} \lambda_{\boldsymbol{\rho},l,m,t}^{\overline{\textrm{C3}}} q_{f(\boldsymbol{\rho}),l,m} P_{m}^{\max}$, and
\[
\begin{alignedat}{1}
& f_{2}\left(\mathbf{D}_{t},\mathbf{s}_{t};\boldsymbol{\lambda}_{t}\right)  = f_{\textrm{Pen}}\left(\mathbf{s}_{t}\right) + \\
& \sum\limits _{\boldsymbol{\rho},l}\mathrm{tr}\Big[\sum\limits _{m}\lambda_{\boldsymbol{\rho},l,m,t}^{\overline{\textrm{C3}}}\boldsymbol{\Lambda}_{m}+\big(1+\sum\limits _{j}\lambda_{\boldsymbol{\rho},l,j,t}^{\overline{\textrm{C8}}}\big)\mathbf{I}_{N}\Big]\mathbf{W}_{\boldsymbol{\rho},l,t}\\
 & +\sum\limits_{\boldsymbol{\rho},l,j}\big(\lambda_{\boldsymbol{\rho},l,j,t}^{\overline{\textrm{C9}}} -\lambda_{\boldsymbol{\rho},l,j,t}^{\overline{\textrm{C8}}}\big)\mathrm{tr}\left(\overline{\mathbf{W}}_{\boldsymbol{\rho},l,j,t}\right)-\sum\limits_{\boldsymbol{\rho},l,j}\lambda_{\boldsymbol{\rho},l,j,t}^{\overline{\textrm{C9}}}P_{\max}\\
 & -\sum\limits _{\boldsymbol{\rho},l,m}\lambda_{\boldsymbol{\rho},l,m,t}^{\overline{\textrm{C3}}} s_{\boldsymbol{\rho},l,m,t}^{\textrm{C3}}+\sum\limits _{\boldsymbol{\rho},l,j}\big(\lambda_{\boldsymbol{\rho},l,j,t}^{\overline{\textrm{C8}}} s_{\boldsymbol{\rho},l,j,t}^{\textrm{C8}}-\lambda_{\boldsymbol{\rho},l,j,t}^{\overline{\textrm{C9}}} s_{\boldsymbol{\rho},l,j,t}^{\textrm{C9}}\big).
\end{alignedat}
\]
$\mathcal{L}_{\mathbf{q}}\left(\mathbf{D}_{t},\mathbf{s}_{t};\boldsymbol{\lambda}_{t}\right)$ is separable w.r.t. $\left\{ \mathbf{q}\right\} $ and $\left\{ \mathbf{D}_{t},\mathbf{s}_{t}\right\}$. Since, for given $\mathbf{q}\in\mathcal{Q}$, problem \eqref{eq:Primal} is  convex and fulfills Slater's condition, the following result holds due to strong duality: 
\vspace{-.2cm}
\begin{equation}
\begin{alignedat}{1}\nu_{t}\left(\mathbf{q}\right) & =\max_{\boldsymbol{\lambda}_{t}\succeq\mathbf{0}\,}\min_{\mathbf{D}_{t}\in\mathcal{D}_{t},\,\mathbf{s}_{t}\succeq \mathbf{0}}\mathcal{L}_{\mathbf{q}}\left(\mathbf{D}_{t},\mathbf{s}_{t};\boldsymbol{\lambda}_{t}\right),\quad\forall\mathbf{q}\in\mathcal{Q}.\end{alignedat}
\label{eq:duality}
\end{equation}
Consequently, the master problem is reformulated as 
\vspace{-.2cm}
\begin{align}
\minimize_{\mathbf{q}\in\mathcal{Q},\alpha}\quad & \alpha\label{eq:Master-2}\\
\st\quad & \alpha\ge\sum\nolimits _{t\in\mathcal{T}_{0}}\xi_{t}\left(\mathbf{q};\boldsymbol{\lambda}_{t}\right),\quad\forall\boldsymbol{\lambda}_{t}\succeq\mathbf{0},\nonumber 
\end{align}
where $\xi_{t}\left(\mathbf{q};\boldsymbol{\lambda}_{t}\right) \triangleq \min_{\mathbf{D}_{t}\in\mathcal{D}_{t},\,\mathbf{s}_{t} \succeq \mathbf{0}} \mathcal{L}_{\mathbf{q}} \left(\mathbf{D}_{t},\mathbf{s}_{t}; \boldsymbol{\lambda}_{t}\right)$. Although problem \eqref{eq:Master-2} still contains an infinite number of constraints (w.r.t. $\boldsymbol{\lambda}_{t}$) and undetermined functions $\xi_{t}(\cdot;\cdot)$, it is readily solvable by an iterative relaxation method as will be explained in the following.

\subsubsection{\label{sub:Iter-Opt-Sol}Optimal Iterative Solution} 

The proposed iterative algorithm is given in Algorithm \ref{alg1}. Let $k$ be the iteration index. We start from one constraint at $k=1$, which defines a cutting plane (also referred as an optimality cut \cite{Floudas1995MINLP}). Then, the number of constraints/cuts are increased sequentially as the iteration proceeds. Specifically, for given dual variables $\boldsymbol{\lambda}_{t}^{j}$, $j=1,\ldots k-1$, the following master problem is solved in iteration $k$,
\vspace{-.25cm}
\begin{align}
\minimize_{\mathbf{q}\in\mathcal{Q},\alpha}\quad & \alpha\label{eq:RM}\\
\st \quad & \alpha\ge\sum\nolimits _{t\in\mathcal{T}_{0}}\xi_{t}\big(\mathbf{q};\boldsymbol{\lambda}_{t}^{j}\big),\quad j=1,\ldots,k-1.\nonumber 
\end{align}
Problem \eqref{eq:RM} is a relaxation of problem \eqref{eq:Master-2}. Due to the enlarged feasible set, the optimal value of problem \eqref{eq:RM} gives a lower bound on that of problem \eqref{eq:Master-2}. The relaxation solution, denoted by $(\mathbf{q}^{k},\alpha^{k})$, is optimal for problem \eqref{eq:Master-2} if it is feasible for problem \eqref{eq:Master-2}. Otherwise, we add another  optimality cut to the feasible set of \eqref{eq:RM} in the next iteration to tighten the relaxation. As this process continues, we obtain a non-decreasing sequence of lower bounds until the relaxed solution becomes feasible, i.e., solves problem \eqref{eq:Master-2} optimally, or until the problem is known to be infeasible.

\begin{algorithm}[t]
\protect\small{\protect\caption{\textcolor{black}{Optimal iterative algorithm for solving P1 and P0}}
\label{alg1} 
\begin{algorithmic}[1] 

\STATE \textbf{Initialization}: Given $\mathbf{q}^{0}\leftarrow\mathbf{0}$. Solve the SDP subproblem \eqref{eq:Primal} for given $\mathbf{q}^{0}$ and determine $\mathbf{D}_{t}^{1},\mathbf{s}_{t}^{1},\boldsymbol{\lambda}_{t}^{1}$; set tolerance $\varepsilon\ge0$, $UB\leftarrow\nu(\mathbf{q}^{0})$, $LB\longleftarrow\infty$, $k\leftarrow1$; 

\WHILE{($UB>LB+\varepsilon$)} \label{line2}

\STATE Solve the relaxed master problem \eqref{eq:RM} for given $\mathbf{D}_{t}^{k}$, $\mathbf{s}_{t}^{k}$, $\boldsymbol{\lambda}_{t}^{k}$ and determine the solution $(\mathbf{q}^{k},\,\alpha^{k})$; \label{line3}

\STATE Update lower bound and solution: $LB \leftarrow \alpha^{k}$, $\mathbf{q}^{*} \leftarrow \mathbf{q}^{k}$; 

\STATE Solve SDP subproblem \eqref{eq:Primal} for given $\mathbf{q}^{k}$ and determine the primal and the dual solutions $\mathbf{D}_{t}^{k+1}$, $\mathbf{s}_{t}^{k+1}$, $\boldsymbol{\lambda}_{t}^{k+1}$;\label{line5}

\IF{($\nu(\mathbf{q}^{k})=+\infty$, i.e., \eqref{eq:Primal} is infeasible, \textbf{OR} $\nu(\mathbf{q}^{k})\le\alpha^{k}+\varepsilon$)}

\STATE Set $\mathbf{D}_{t}^{*}\leftarrow\mathbf{D}_{t}^{k+1}$, $\mathbf{s}_{t}^{*} \leftarrow \mathbf{s}_{t}^{k+1}$ and exit the while loop;

\ELSIF{($\nu(\mathbf{q}^{k})<UB$)}

\STATE Update upper bound and solution: $UB\leftarrow\nu(\mathbf{q}^{k})$, $\mathbf{D}_{t}^{*} \leftarrow \mathbf{D}_{t}^{k+1}$, $\mathbf{s}_{t}^{*}\leftarrow\mathbf{s}_{t}^{k+1}$; \label{line9}

\ENDIF \label{line10}

\STATE Update iteration index: $k\leftarrow k+1$;

\ENDWHILE 

\IF{($\mathbf{s}_{t}^{*}=\mathbf{0}$)}

\STATE Return the optimal solutions $\mathbf{q}^{*}$ and $\mathbf{D}_{t}^{*}$;

\ELSE 

\STATE Return the infeasible problem P0/P1.

\ENDIF 

\end{algorithmic} }
\end{algorithm}

Two remarks regarding Algorithm 1 are in order. First, as can be observed in Algorithm~\ref{alg1} (lines~\ref{line5}--\ref{line10}), the feasibility or optimality of $\mathbf{q}^{k}$ is verified by solving SDP subproblem \eqref{eq:Primal}. This is because, if $\mathbf{q}^{k}$ is optimal, we know that solving problem \eqref{eq:Primal} for $\mathbf{q} = \mathbf{q}^{k}$ in line~\ref{line5} will return the optimal value of $\alpha^{k}$, i.e., $\nu(\mathbf{q}^{k})=\alpha^{k}$, owing to the strong duality of problem \eqref{eq:Primal}. Otherwise, $\nu(\mathbf{q}^{k})$ gives an upper bound on the optimal value, and thus, $\nu(\mathbf{q}^{k}) \ge \alpha^{k}$. By keeping the lowest upper bound obtained so far, i.e., $UB \leftarrow \min \left\{ UB,\,\nu(\mathbf{q}^{k}) \right\}$ (cf. line~\ref{line9}), the optimality condition is satisfied when the gap between $UB$ and the lower bound vanishes.

Second, for computational convenience, the values of $\boldsymbol{\lambda}_{t}^{k}$ in iteration $k$ can be intelligently chosen as the optimal dual solutions of problem \eqref{eq:duality} or \eqref{eq:Primal} for  $\mathbf{q}=\mathbf{q}^{k}$ in line~\ref{line5}. In this case, the constraint function $\xi_{t}(\cdot;\cdot)$ can be easily computed as explained in the following proposition.
\begin{prop}
\emph{\label{prop2}Let $(\mathbf{D}_{t}^{k},\,\mathbf{s}_{t}^{k})$ and $\boldsymbol{\lambda}_{t}^{k}$ be the optimal primal and dual solutions of \eqref{eq:Primal} for $\nu (\mathbf{q}^{k})$ 
at iteration $k$, respectively. Then, we have i) $(\mathbf{D}_{t}^{k},\,\mathbf{s}_{t}^{k})$ also solves the minimization problem in the optimality cut of iteration $k+1$ (cf. \eqref{eq:RM}), i.e., 
$(\mathbf{D}_{t}^{k},\,\mathbf{s}_{t}^{k}) \in {\arg\min}_{\mathbf{D}_{t}\in\mathcal{D}_{t},\,\mathbf{s}_{t}\ge\mathbf{0}} \; \mathcal{L}_{\mathbf{q}}(\mathbf{D}_{t},\mathbf{s}_{t};\boldsymbol{\lambda}_{t}^{k}).$
}

\emph{ii) By choosing $\boldsymbol{\lambda}_t = \boldsymbol{\lambda}_{t}^{k}$, function $\xi_{t}\left(\mathbf{q};\boldsymbol{\lambda}_{t}^{k} \right)$ reduces to an affine function given by
\begin{alignat}{1}
\xi_{t}\left(\mathbf{q};\boldsymbol{\lambda}_{t}^{k}\right)= & \sum\limits_{\boldsymbol{\rho},l,j} \!\! \big(\lambda_{\boldsymbol{\rho},l,j,t}^{\overline{\textrm{C9}},k} - \lambda_{\boldsymbol{\rho},l,j,t}^{\overline{\textrm{C8}},k} \big) P_{\max} \!\! \left(q_{f(\boldsymbol{\rho}),l,j} - q_{f(\boldsymbol{\rho}),l,j}^{k}\right) \nonumber \\
 & -\sum\limits_{\boldsymbol{\rho},l,m}\lambda_{\boldsymbol{\rho},l,m,t}^{\overline{\textrm{C3}},k}P_{m}^{\max}\left(q_{f(\boldsymbol{\rho}),l,m}-q_{f(\boldsymbol{\rho}),l,m}^{k}\right) \nonumber \\ 
& +U_{\mathrm{TP}}\left(\mathbf{q}^{k},\mathbf{D}_{t}^{k}\right) 
+f_{\textrm{Pen}}\left(\mathbf{s}_{t}^{k}\right).
\label{eq:psi}
\end{alignat} }
\end{prop}
\begin{IEEEproof}
Please refer to Appendix~\ref{proof2}.
\end{IEEEproof}
Based on Proposition~\ref{prop2}, the relaxed master problem \eqref{eq:RM} is an MILP and can be solved optimally, e.g., using the numerical solver MOSEK \cite{MOSEK}. Similar to the conventional GBD method, Algorithm~\ref{alg1} converges in a finite number of iterations as shown in Proposition~\ref{prop3}. The obtained solution is globally optimal for problem P1. In general, the solution of P1 gives a lower bound for problem P0. However, by inspecting the rank of the SDP solution of problem P1, we can further show that the SDP relaxation is tight. 
\begin{prop}
\emph{\label{prop3} Algorithm \ref{alg1} converges in a finite number iterations. Moreover, assuming that the channel vectors $\widehat{\mathbf{h}}_{\boldsymbol{\rho},t}$, $\boldsymbol{\rho}\in\mathcal{S}$, can be modeled as statistically independent random vectors,  problems P1 and P0 are equivalent in the sense that whenever P0 is feasible, the solution of P1 is also (globally) optimal for P0 with probability one, and the optimal beamformer is given by the principal eigenvector of $\mathbf{W}_{\boldsymbol{\rho},l,t}$.} 
\end{prop}
\begin{IEEEproof}
Please refer to Appendix~\ref{proof3}. 
\end{IEEEproof}

Due to perturbation, only optimality cuts, cf. \eqref{eq:RM}, need to be generated by Algorithm \ref{alg1} in each iteration, cf. Remark~\ref{rem3}. This is different from the classical GBD algorithm \cite[Section 6.3]{Floudas1995MINLP} where feasibility cuts are also required to exclude infeasible solutions during intermediate iterations. Since the optimality cuts can successively improve the lower bounds, Algorithm~\ref{alg1} is expected to converge faster than the classic GBD algorithm if P1 is feasible. On the other hand, even if P1 is infeasible, the perturbed problem is generally feasible. Then, optimality cuts can be still generated to iteratively improve the solutions  and reduce the required number of iterations with a high probability.

\subsubsection{Computational Complexity}
Assume that the interior-point method \cite{Ye2011interior,Boyd2004Convex} is applied to solve the SDP subproblems in each iteration of the GBD algorithm. The computational complexity of solving each SDP subproblem w.r.t. the number of UEs, $K$, the number of BSs, $M$, the number of BS antennas, $N$, and the number of SVC layers, $L$, can be approximated as \cite[Theorem 3.12]{polik10interior}
\begin{alignat}{1}
\Theta^{\mathrm{sdp}} & =\mathcal{O}\Big( \underset{\textrm{Complexity per iteration}}{\underbrace{ \left(\left(MKL\right)^{4}\left(N^{3}+N^{2}+2\right)+\left(MKL\right)^{3}\right) }}  \nonumber \\
 & \qquad\qquad \times \underset{\textrm{Number of iterations}}{\underbrace{ \sqrt{MKLN}\log\left(\epsilon^{-1}\right)}} \Big)\nonumber \\
 &  = \mathcal{O}\left(\left(MKL\right)^{4.5}N^{3.5}\log\left(\epsilon^{-1}\right)\right),
\label{complexity}
\end{alignat}
where $\epsilon \!>\! 0$ is the solution accuracy specified by the numerical solver and $\mathcal{O}(\cdot)$ is the big-$O$  notation. Although the SDP subproblems can be solved in polynomial time in line~\ref{line5}, cf. \eqref{complexity}, the overall computational complexity of Algorithm~\ref{alg1} grows non-polynomially with the size of problem P0. This is because the MILP solver in line~\ref{line3} may incur an exponential-time computational complexity, $\mathcal{O}(2^{FLM})$, in the worst case \cite{Floudas1995MINLP}, even though the likelihood that the worst case occurs is low due to the employed perturbation, cf. Remark~\ref{rem3}. Thus, only offline cache optimization may be feasible in practical implementations. 

\vspace{-0.5cm}
\subsection{\label{sec-iv-b}Suboptimal Caching Scheme}
For systems with limited computing resources, Algorithm~\ref{alg1} may not be applicable due to its worst-case exponential-time computational complexity. Instead, polynomial-time suboptimal schemes facilitating a better trade-off between system performance and computational complexity may be preferable. Based on Proposition \ref{prop3}, P0 can be solved via its equivalent convex MINLP, P1. As is also evident from \eqref{eq:Primal}, for given  $\mathbf{q}$, P1 reduces to an SDP and can be solved optimally in polynomial time, cf. \eqref{complexity}. Therefore, by additionally adjusting $\mathbf{q}$ via a greedy iterative search, we obtain the low-complexity suboptimal scheme in Algorithm~\ref{alg2}.

\begin{algorithm}[t]
\protect\small{\protect\caption{Suboptimal iterative algorithm for solving P1 and P0}

\label{alg2} \begin{algorithmic}[1] 

\STATE \textbf{Initialization}: Given $\mathbf{q}_{m}^{1} \leftarrow \mathbf{0}$, $\forall m\in\mathcal{M}$; $k\leftarrow1$;

\WHILE{$\mathcal{I}^{k}\neq\emptyset$}

\FOR{\textbf{each} $i\in\mathcal{I}^{k}$}

\STATE Solve SDP subproblem \eqref{eq:Primal} for each given $\left\{ \mathbf{q}_{m}\right\} $ satisfying $\mathbf{q}_{i}\in\mathcal{Q}_{i}^{k}\cap\mathcal{Q}$; 

\STATE Determine $\mathbf{q}_{i}^{k+1}$ in \eqref{eq:greedy};

\ENDFOR

\STATE $k\leftarrow k+1$.

\ENDWHILE

\end{algorithmic} }
\end{algorithm}

Let $\mathcal{F}_{\mathcal{S}}$ and $\mathbf{q}_{m}$ be the set of files requested by  $\mathcal{S}$ (the set of requests) and the caching vector at BS $m$, respectively, where $\mathcal{F}_{\mathcal{S}} \subseteq \mathcal{F}$. We define 
\vspace{-.2cm}
\begin{equation}
\mathcal{Q}_{m}^{k}\triangleq\Big\{ \mathbf{q}_{m}\in\left\{ 0,1\right\} ^{\left|\mathcal{F}_{\mathcal{S}}\right|\times L}\mid \left\Vert \mathbf{q}_{m}-\mathbf{q}_{m}^{k} \right\Vert_2^2 \le1\Big\} 
\label{subopt1}
\end{equation}
as the set of binary vectors within a distance of one from $\mathbf{q}_{m}^{k}$. Besides, $\mathcal{I}^{k}\triangleq \big\{ m \in \mathcal{M}\mid\left|\mathcal{Q}_{m}^{k} \cap \mathcal{Q}\right|>1\big\} $ defines the set of BS indices where $\mathcal{Q}_{m}^{k}$ and $\mathcal{Q}$ have non-unique intersection points. During iteration $k$, the vector in set $\mathcal{Q}_{i}^{k}\cap\mathcal{Q}$ that minimizes the objective value of primal problem \eqref{eq:Primal} is chosen as the new caching vector at BS $i\in\mathcal{I}^{k}$, i.e., 
\begin{equation}
\mathbf{q}_{i}^{k+1} = \underset{\mathbf{q}_{i}\in\mathcal{Q}_{i}^{k}\cap\mathcal{Q}}{\arg\min} \quad \nu \left(\mathbf{q}_{1},\ldots,\mathbf{q}_{M}\right).
\label{eq:greedy}
\end{equation}
That is, the cache vectors $\mathbf{q}_{i}^{k+1}$, within a distance of one from $\mathbf{q}_{i}^{k}$, are iteratively updated to successively reduce the objective value. The iteration continues until $\mathcal{Q}_{i}^{k}\cap\mathcal{Q}$ becomes unique, i.e., no further reduction in the objective function is possible, which yields the solution. Hence, the number of problem instances of \eqref{eq:Primal} to be solved is bounded by $ML^{2}\left|\mathcal{F}_{\mathcal{S}}\right|^{2}$. Consequently, the overall computational complexity of Algorithm \ref{alg2} is approximated as $\mathcal{O}\left(ML^{2}\left|\mathcal{F}_{\mathcal{S}}\right|^{2}\Theta^{\mathrm{sdp}}\right)$, which grows only polynomially with the problem size. 
\vspace{-0.2cm}
\begin{rem}
By adopting the greedy heuristic, searching over the non-convex set $\mathcal{Q}$ of P1 can be done in polynomial time. The obtained solution is ensured to be feasible for P1. Moreover, it is often close-to-optimal due to the iterative minimization in \eqref{eq:greedy} \cite{Korte2002combinatorial}, as will be shown in Section~\ref{sec:Simulation-Results}.  
\end{rem}

\vspace{-.5cm}
\subsection{\label{sec-iv-c}Optimal Delivery Solution}
By applying the same transformation techniques as for problem P0 in Section~\ref{sec-iv-a} and relaxing the rank constraint, cf. C11, problem Q0 can be reformulated as an SDP, which is equivalent to problem \eqref{eq:Primal}. 
Since, based on Proposition~\ref{prop3}, the solution of problem \eqref{eq:Primal} fulfills rank constraint C11 with probability one,  delivery problem Q0 can be solved optimally via SDP subproblem \eqref{eq:Primal}, as stated in Corollary \ref{cor1}. 
\begin{cor}
\emph{\label{cor1} SDP subproblem \eqref{eq:Primal} (for the respective instantaneous CSI) and the delivery optimization problem Q0 are equivalent in the sense that the solution of \eqref{eq:Primal} is also optimal for Q0 whenever Q0 is feasible. } 
\end{cor}
\vspace{-.2cm}  
\begin{IEEEproof}
The proof is similar to that of Propositions \ref{prop1} and \ref{prop3} and is omitted for brevity.
\end{IEEEproof}
Therefore, delivery optimization problem Q0 can be solved in polynomial time with computational complexity $\Theta^{\mathrm{sdp}}$, cf. \eqref{complexity}, which is desirable for online implementation \cite{palomar2010convex}. Moreover, delivery optimization incurs a signaling overhead of $\mathcal{O}(MKLN)$ for collecting the CSI at the macro BS and distributing the optimization results to the small cell BSs.

\vspace{-.2cm}
\begin{rem}
Although the total number of BSs in dense small cell networks may be large \cite{Ge16UDN}, the coverage areas of most small cell BSs will not overlap. Therefore, the proposed caching and delivery algorithms may be applied to several small groups of small cell BSs with overlapping coverage areas rather than jointly to all small cell BSs. This considerably reduces the computational complexity and signaling overhead.
\end{rem}

\vspace{-.2cm}
\begin{rem}
The proposed caching and delivery optimization framework can be extended to integrate (centralized and decentralized) coded caching at the end users \cite{Maddah14:Centralized,Maddah15:Decentralized,Zhang15:Coded,Naderializadeh17,Zewail:ISIT17}. For example,  multicast codewords can be cooperatively transmitted by a subset of the BSs if these BSs have cached the subfiles required for coded multicast transmission. Such a design reaps the performance gains of both coded caching and cache-enabled cooperative MIMO transmission. However, the resulting cache optimization problem would involve a large number of binary caching variables, which have to be defined per subfile, and SDP relaxation of the delivery optimization problem may not yield the optimal solution anymore \cite{Luo06Multicast}. Hence, extending the proposed framework to coded caching is an interesting topic for future research.
\end{rem} 

\section{\label{sec:Simulation-Results}Simulation Results}
In this section, we evaluate the performance of the proposed optimal and suboptimal schemes. Consider a cell of radius $R_{1}=1$~km, where the macro BS is located at the center of the cell and three small cell BSs are uniformly distributed within the cell. The number of untrusted small cell BSs is set to $M_{u} \!\triangleq\! \left| \mathcal{M_U} \right| =1$, unless stated otherwise. To gain insight, in Figs.~2--5, we consider a small network with only three small cell BSs. A larger network is considered in Fig.~6. The macro BS is equipped with $N_{0}=6$ antennas while each small cell BS has $N_{m}=2$ antennas. We assume that $F \!=\! 10$ video files, each of duration $45$~minutes and size $500$~MB (Bytes), are delivered to $K \!=\! 6$ single-antenna UEs. Each user requests the files independent of the other users. Let $\boldsymbol{\theta}=\left[\theta_{1},\ldots,\theta_{F}\right]$ be the probability distribution of the requests for different files. We set $\boldsymbol{\theta}$ according to a Zipf distribution with $\gamma=1.1$. In particular, assuming that file $f \!\in\! \mathcal{F}$ is the $\sigma_f$th most popular file for the UEs, the probability of file $f \!\in\! \mathcal{F}$ being requested is given by $\theta_{f} \!=\! \frac{1}{\sigma_f^{\gamma}}/\sum_{f\in\mathcal{F}}\frac{1}{\sigma_f^{\gamma}}$ \cite{Breslau99Zipf}. We adopt an SVC codec with $L \!=\! 2$. That is, each video file is encoded into a base-layer subfile ($l=0$) and an enhancement-layer subfile ($l=1$), each of size $V_{f,l}=250$~MB. The minimum streaming rate and the secrecy rate threshold for the base-layer subfiles are $R_{\boldsymbol{\rho},0}^{\textrm{req}} \!=\! 825$~kbps and $R_{\boldsymbol{\rho},0}^{\mathrm{tol}} \!=\! 0.1R_{\boldsymbol{\rho},0}^{\textrm{req}} \!=\! 82.5$~kbps, respectively. Therefore, if problem Q0 is feasible, a secrecy streaming rate of $R_{\boldsymbol{\rho},0,t}^{\textrm{sec}}=742.5$~kbps can be guaranteed for secure and uninterrupted video streaming for each user as $R_{\boldsymbol{\rho},0,t}^{\textrm{sec}} \!\ge\! 250 \!\times\! 8 \!\times\! 10^{6}/(45 \!\times\! 60) \!=\! 741$~kbps. The streaming rate of the enhancement-layer subfiles is $R_{\boldsymbol{\rho},1}^{\textrm{req}} \!=\! 2R_{\boldsymbol{\rho},0,t}^{\textrm{sec}} \!=\! 1.5$~Mbps. The users are randomly distributed in the system. Based on the locations of the BSs and users, the path loss is calculated using the 3GPP model for the ``urban macro non-line-of-sight'' scenario \cite{IEEE3GPP:TR36814}. The small-scale fading coefficients are independent and identically distributed (i.i.d.) Rayleigh random variables. We employ Euclidean spheres for modeling the uncertainty regions $\boldsymbol{\Omega}_{\boldsymbol{\rho},t}$ and $\boldsymbol{\Omega}_{j,t}$ by setting $\boldsymbol{\Xi}_{\boldsymbol{\rho}} = \boldsymbol{\Xi}_{j} = \mathbf{I}_{N}$. Meanwhile, we define the maximum normalized channel estimation error variances of $\mathbf{h}_{\boldsymbol{\rho},t}$ and $\mathbf{G}_{j,t}$ as $\sigma_{\boldsymbol{\rho}}^{2} \!=\! \frac{\varepsilon_{\boldsymbol{\rho}}^{2}}{\left\Vert \mathbf{h}_{\boldsymbol{\rho},t}\right\Vert _{2}^{2}}$ and $\sigma_{j}^{2} \!=\! \frac{\varepsilon_{j}^{2}}{\left\Vert \mathbf{G}_{j,t}\right\Vert _{F}^{2}}$, respectively. Unless otherwise specified, we assume $\sigma_{\boldsymbol{\rho}}^{2}= 0.01$, $\boldsymbol{\rho}\in\mathcal{S}$ and $\sigma_{j}^{2}= 0.05$, $j \in \mathcal{M_{U}}$. All other relevant system parameters are given in Table~\ref{tab1}. 
\begin{table}
\centering \protect \caption{\label{tab1}Simulation Parameters.}
\small{
\begin{tabular}{|c|c|}
\hline 
{ Parameters} & { Settings} \tabularnewline
\hline 
\hline 
{ System bandwidth} & { 5 MHz} \tabularnewline
\hline 
{ Duration of time slot} & { 10 ms} \tabularnewline
\hline 
{ Duration of delivery period} & { 45 mins } \tabularnewline
\hline 
{ Macro BS transmit power} & { $P_{0}^{\max}$= 46 dBm } \tabularnewline
\hline 
{ Small BS transmit power} & { $P_{m}^{\max}$= 39 dBm} \tabularnewline
\hline 
{ Noise power density} & { $-$172.6 dBm/Hz} \tabularnewline
\hline 
{ Cache capacity at macro BS} & { $C_{0}^{\max}$= 1000 MB} \tabularnewline
\hline 
\end{tabular} }
\end{table}

For comparison, we consider two heuristic caching schemes and a non-cooperative and a non-robust delivery scheme as baselines: 
\begin{itemize}
\item Baseline 1 (Random caching): The video (sub)files are randomly cached until the cache capacity is reached.

\item Baseline 2 (Preference based caching): The most popular (sub)files are cached. In trusted BSs, since the base-layer subfiles are more important, they are cached with higher priority than the enhancement-layer subfiles of the same video file. For Baselines 1 and 2, the optimal delivery decisions are obtained by solving problem Q0.

\item Baseline 3 (No cooperation with untrusted BSs): No video files are cached at the untrusted BSs, which act as pure eavesdroppers. Hence, the untrusted BSs are not allowed to cooperate for delivery of the video files. This approach is adopted in state-of-the-art cellular networks. The optimal caching and delivery decisions are obtained from problems P0 and Q0, respectively, with $C_{m}^{\max}=0$, $\forall m\in\mathcal{M_{U}}$. 

\item Baseline 4 (Non-robust transmission): Different from the proposed schemes, the macro BS treats the channel estimates $\widehat{\mathbf{h}}_{\boldsymbol{\rho},t}$ and $\widehat{\mathbf{G}}_{j,t}$ as accurate. The optimal caching and delivery decisions are obtained by solving problems P0 and Q0, respectively, after setting $\mathbf{h}_{\boldsymbol{\rho},t}=\widehat{\mathbf{h}}_{\boldsymbol{\rho},t}$ and $\mathbf{G}_{j,t}=\widehat{\mathbf{G}}_{j,t}$.
\end{itemize}

\begin{figure}[t]
\centering
\subfloat[]{\includegraphics[width=3.0in]{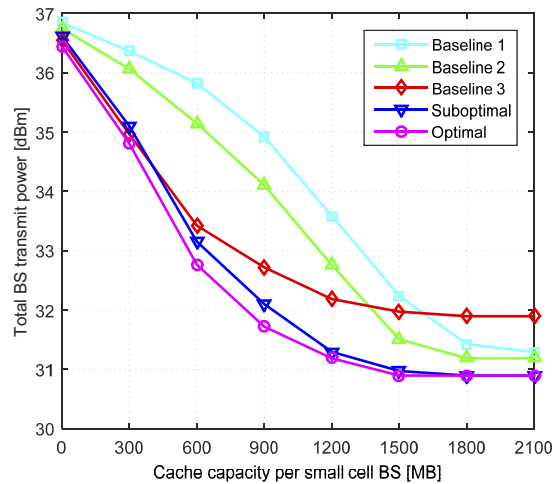} \label{fig2a}} \;
\subfloat[]{\includegraphics[width=3.0in]{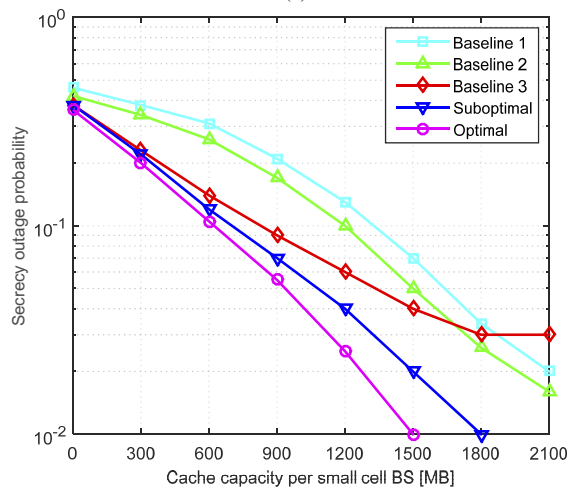} \label{fig2b}} 
\protect\protect\caption{\label{fig2}\small (a) Total BS transmit power and (b) secrecy outage probability versus cache capacity for different caching and delivery schemes.}
\vspace{-.4cm}
\end{figure}

Figs.~\ref{fig2a} and \ref{fig2b} illustrate the performance of the considered caching and delivery schemes as functions of the cache capacity. Herein, the system performance is evaluated during the online delivery of the video files, cf. problem Q0. The secrecy outage probability, defined as $p_{\mathrm{out}} \triangleq \Pr \big(\sum_{\boldsymbol{\rho}} R_{\boldsymbol{\rho},0,t}^{\textrm{sec}} < \sum_{\boldsymbol{\rho}} \left[R_{\boldsymbol{\rho},0}^{\textrm{req}} - R_{\boldsymbol{\rho},0}^{\mathrm{tol}} \right]^{+}\big)$, characterizes the likelihood that problem Q0 is infeasible because either the QoS constraint C6 or the secrecy constraint C7 cannot be satisfied. As can be observed from Figs.~\ref{fig2a} and \ref{fig2b}, for all considered schemes, a larger cache capacity leads to both a lower total BS transmit power and a smaller secrecy outage probability as larger virtual transmit antenna arrays can be formed among the trusted and untrusted BSs for cooperative beamforming transmission of the base-layer and enhancement-layer subfiles, respectively. There is a non-negligible performance gap between the optimal scheme and Baseline 3, particularly in the high cache capacity regime. This is because the proposed caching scheme can exploit the cache resources of the untrusted helpers for delivering enhancement-layer subfiles while Baseline 3 cannot. The performance gap between the proposed optimal scheme and Baselines 1 and 2 is small for small (large) cache capacities because of insufficient (saturated) BS cooperation. For medium cache capacities, however, the proposed optimal scheme achieves considerable performance gains due to its ability to exploit information regarding the user requests and CSI for cache placement. We note that the proposed suboptimal scheme attains good performance in all regimes despite its low computational cost.

\begin{figure}[t]
\centering
\includegraphics[width=3.0in]{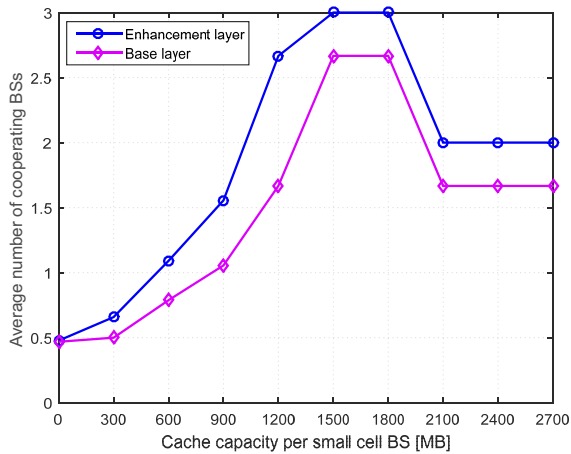} 
\protect\protect\caption{\label{fig2c} \small Average number of cooperating BSs versus cache capacity for transmission of the base-layer and enhancement-layer subfiles when Algorithm~\ref{alg1} is employed for cache optimization.}
\vspace{-.4cm}
\end{figure}

To provide more insight into how the BSs cooperate, Fig.~\ref{fig2c} shows the average numbers of cooperating (small cell and macro) BSs for transmission of the base-layer and enhancement-layer subfiles, denoted as $\overline{N}_{\mathrm{BL}}$ and $\overline{N}_{\mathrm{EL}}$, respectively, if Algorithm~\ref{alg1} is employed for cache optimization. Recall that the proposed caching scheme does not cache base-layer subfiles at untrusted BSs, cf. constraint C1. Consequently, for a given cache capacity, $\overline{N}_{\mathrm{BL}} \le \overline{N}_{\mathrm{EL}}$ holds. Interestingly, the behavior of $\overline{N}_{\mathrm{BL}}$ and $\overline{N}_{\mathrm{EL}}$ is not monotonic as the cache capacity increases. In the small and medium cache capacity regime, $\overline{N}_{\mathrm{BL}}$ and $\overline{N}_{\mathrm{EL}}$ monotonically increase with the cache capacity since the number of subfiles that can be cached at the small cell BSs and the number of BSs that can participate in cooperative transmission increase. For example, $\overline{N}_{\mathrm{BL}}$ and $\overline{N}_{\mathrm{EL}}$ are 0.5 and 0.7 for $300$~MB cache capacity per small cell BS, respectively, and increase to 1.7 and 2.7 for $1200$~MB cache capacity per small cell BS, respectively. 
In the large cache capacity regime, the performance gains saturate as the available DoFs for the transmission of the base-layer and enhancement-layer subfiles saturate, cf. Fig.~\ref{fig2}. However,  $\overline{N}_{\mathrm{BL}}$ and $\overline{N}_{\mathrm{EL}}$  decrease by 1 before reaching the stationary BS cooperation topology. This is because, when the cache-enabled DoFs are sufficient, the optimal caching scheme selects preferred trusted cooperating (small cell and/or macro) BSs, e.g., based on their channel conditions and cache status, instead of exploiting all BSs available for cooperation.

\begin{figure}[t]
\centering
\subfloat[]{\includegraphics[width=3.0in]{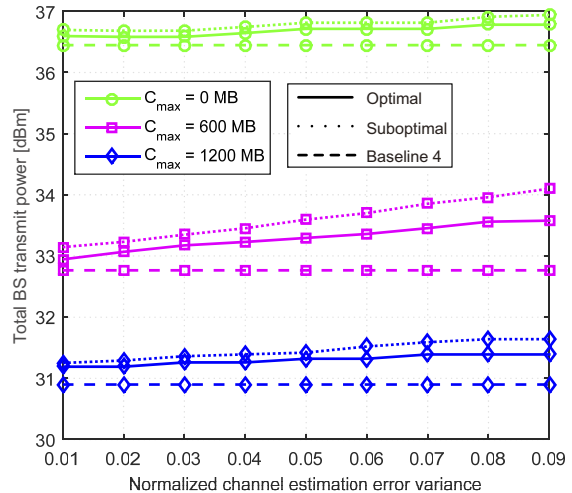} \label{fig3a}} \;
\subfloat[]{\includegraphics[width=3.0in]{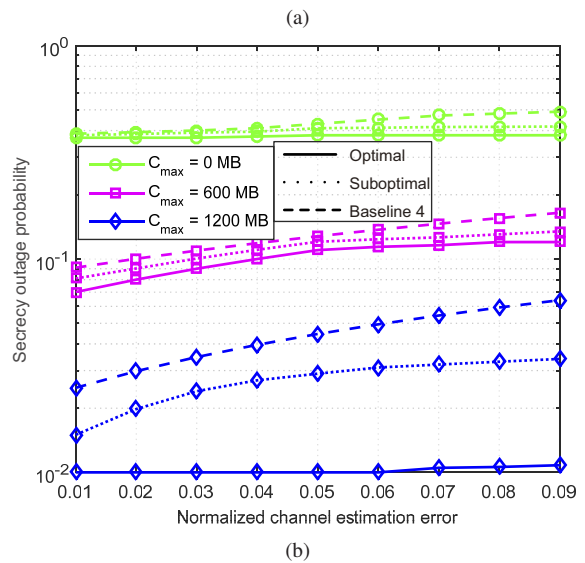} \label{fig3b}}
\protect\protect\caption{\small (a) Total BS transmit power and (b) secrecy outage probability versus normalized channel estimation error variance for the proposed optimal scheme\! (solid~line), the suboptimal scheme\! (dotted~line), and Baseline~4\! (dashed~line).}
\vspace{-.4cm}
\end{figure}

Next, we evaluate the robustness of the proposed schemes w.r.t. channel estimation errors. Figs.~\ref{fig3a} and \ref{fig3b} show the performance of the proposed schemes and Baseline 4 as functions of the normalized channel estimation error variances, $\sigma_{j}^{2}$, where $C_{\max}$ denotes the cache capacity per small cell BS. We observe that, compared to Baseline 4, the proposed schemes achieve a lower secrecy outage probability at the cost of a slightly higher transmit power consumption. Specifically, to achieve robustness in meeting QoS constraint C6 under imperfect CSI, the proposed schemes employ wide beams for transmitting the base-layer subfiles, which may lead to information leakage to the untrusted BSs. Hence, to ensure communication secrecy in constraint C7, the proposed schemes also have to transmit a non-negligible amount of AN to degrade the reception of the untrusted BSs. On the other hand, the wide beams and the interference caused by the AN to the legitimate users, cf. \eqref{leg-rate}, have to be compensated by increasing the transmit power of the beamforming vectors. Therefore, the total transmit power increases as the CSI uncertainty increases. In contrast, by treating the imperfect CSI as perfect in C6, Baseline 4 employs narrow transmit beams to save transmit power but this leads to the highest secrecy outage probability in Fig.~\ref{fig3b}.

\begin{figure}[t]
\centering
\subfloat[]{\includegraphics[width=3.0in]{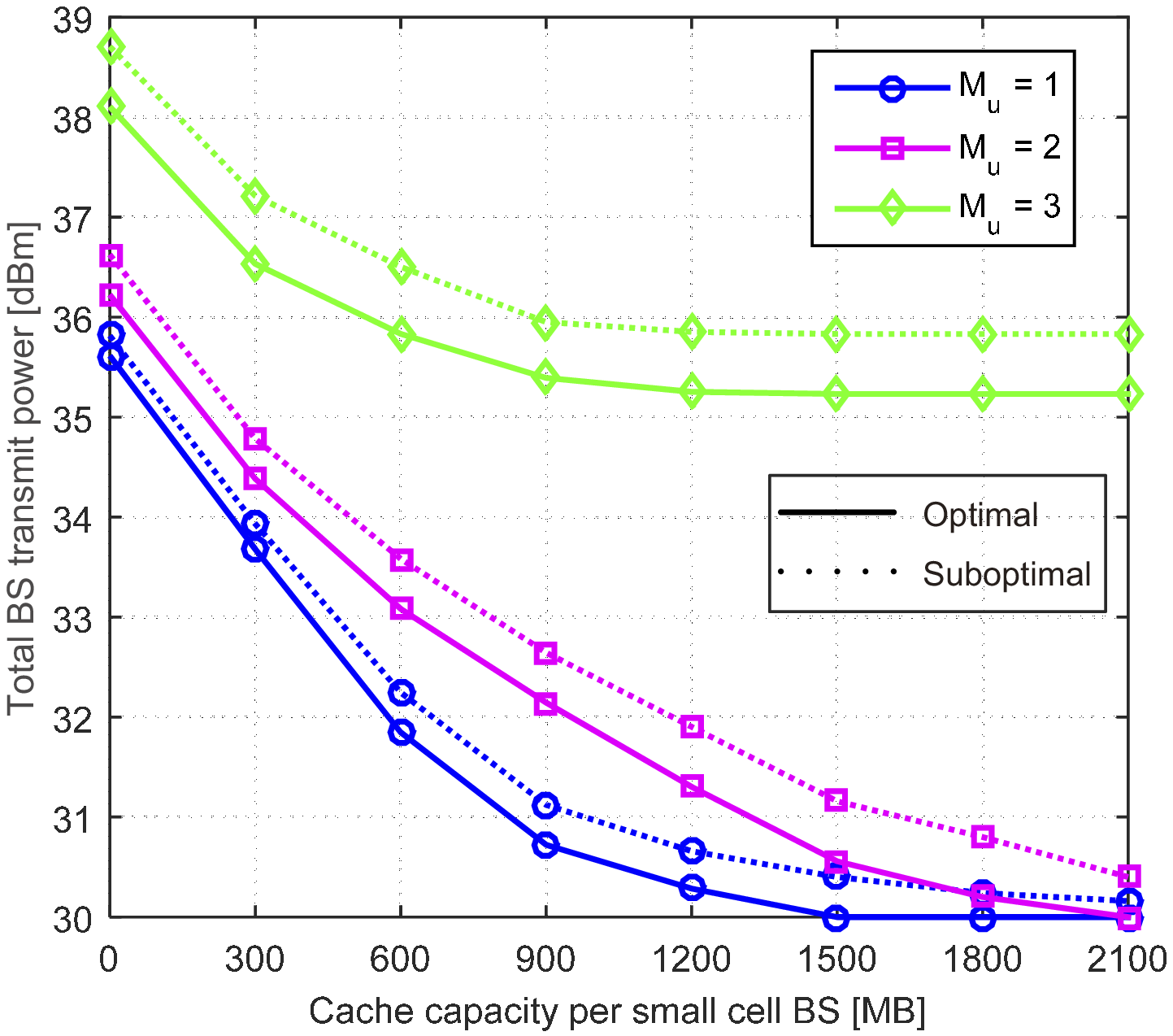} \label{fig4a}} \;
\subfloat[]{\includegraphics[width=3.0in]{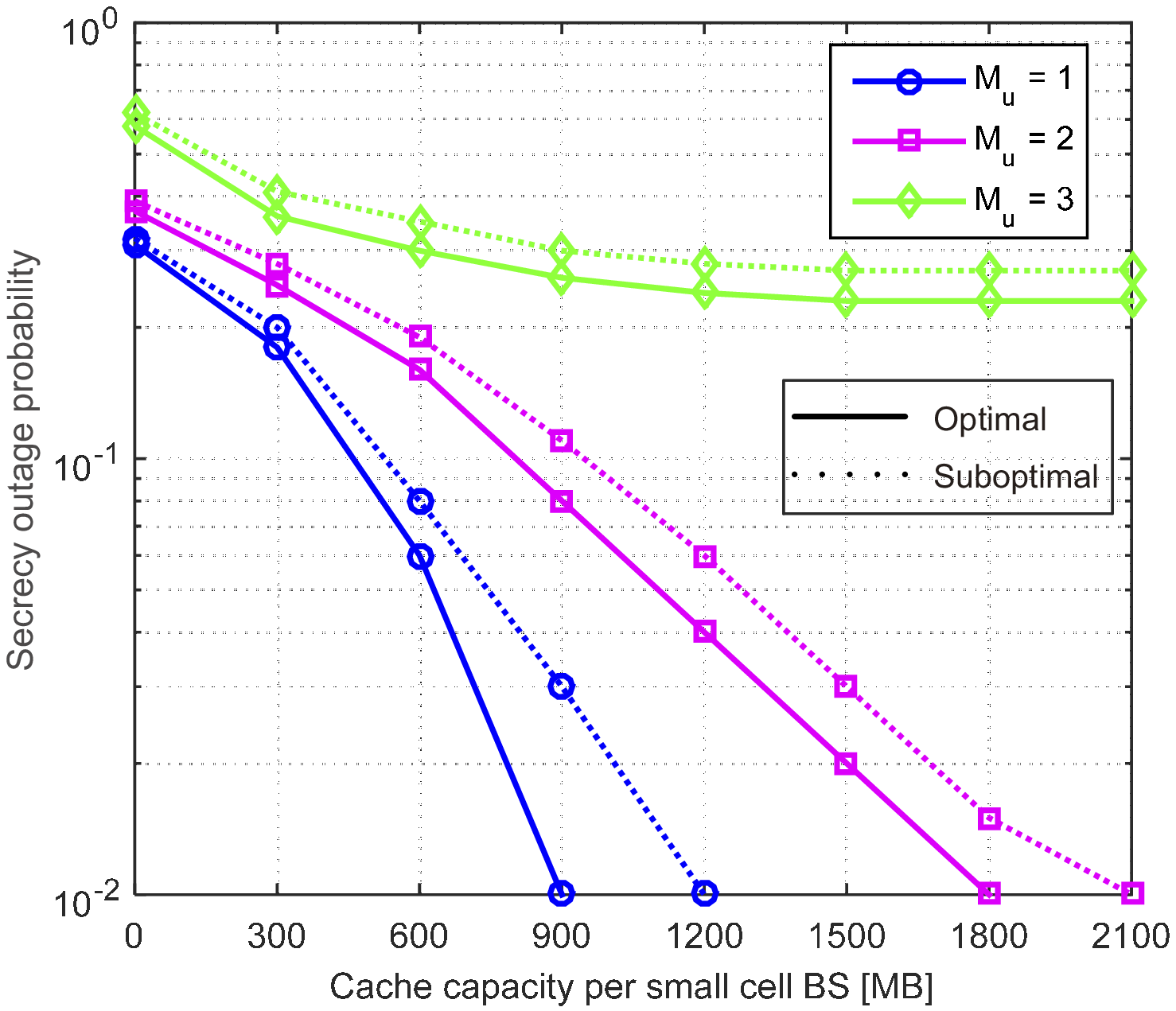} \label{fig4b}}
\protect\protect\caption{\small (a) Total BS transmit power and (b) secrecy outage probability of the proposed optimal (solid line) and suboptimal (dotted line) schemes versus cache capacity for different numbers of untrusted BSs.}
\vspace{-.4cm}
\end{figure}

The impact of the number of trusted and untrusted BSs on cache-enabled secrecy is studied in Figs.~\ref{fig4a} and \ref{fig4b}. Fig.~\ref{fig4a} reveals that the required transmit power increases with the number of untrusted BSs $M_{u}$, if the total number of BSs is kept constant. This is because, as more helpers become untrusted, fewer (trusted) BSs are available for cooperative transmission of the base-layer subfiles and, at the same time, the trusted BSs have to transmit a larger amount of AN to combat the increasing number of potential eavesdroppers. On the other hand, as the base layers are not cached at the untrusted BSs, for a larger $M_u$, more cache capacity can be utilized to transmit the enhancement-layer subfiles. Hence, the transmit power of the optimal/suboptimal scheme is only enlarged moderately when $M_{u}$ increases from $1$ to $2$; in the high cache capacity regime, the increase in transmit power is even negligible. Due to the cooperative transmission of the base-layer and enhancement-layer subfiles, in Fig.~\ref{fig4b}, the secrecy outage probability  monotonically decreases with the cache capacity for $M_{u}\le2$. However, when $M_{u}$ increases from $2$ to $3$, both the transmit power and the secrecy outage probability are increased significantly; particularly, their values saturate at high levels for cache capacities exceeding 600 MB. This is because, for $M_{u}=3$, the total number of antennas equipped at the untrusted BSs equals the total number of antennas equipped at the trusted BSs; hence, the available DoFs for secure transmission of the base-layer subfiles are limited, irrespective of the cache capacities, as the cache of the untrusted small cell BSs can only facilitate the cooperative transmission of enhancement-layer subfiles. Hence, for secure delivery of the base-layer subfiles, the system has to allocate large amounts of power to AN transmission to degrade the reception quality of the untrusted BSs and, at the same time, to the users' signals to mitigate the impact of the interference caused by AN to the legitimate users.

\begin{figure}[t]
\centering
\subfloat[]{\includegraphics[width=3.0in]{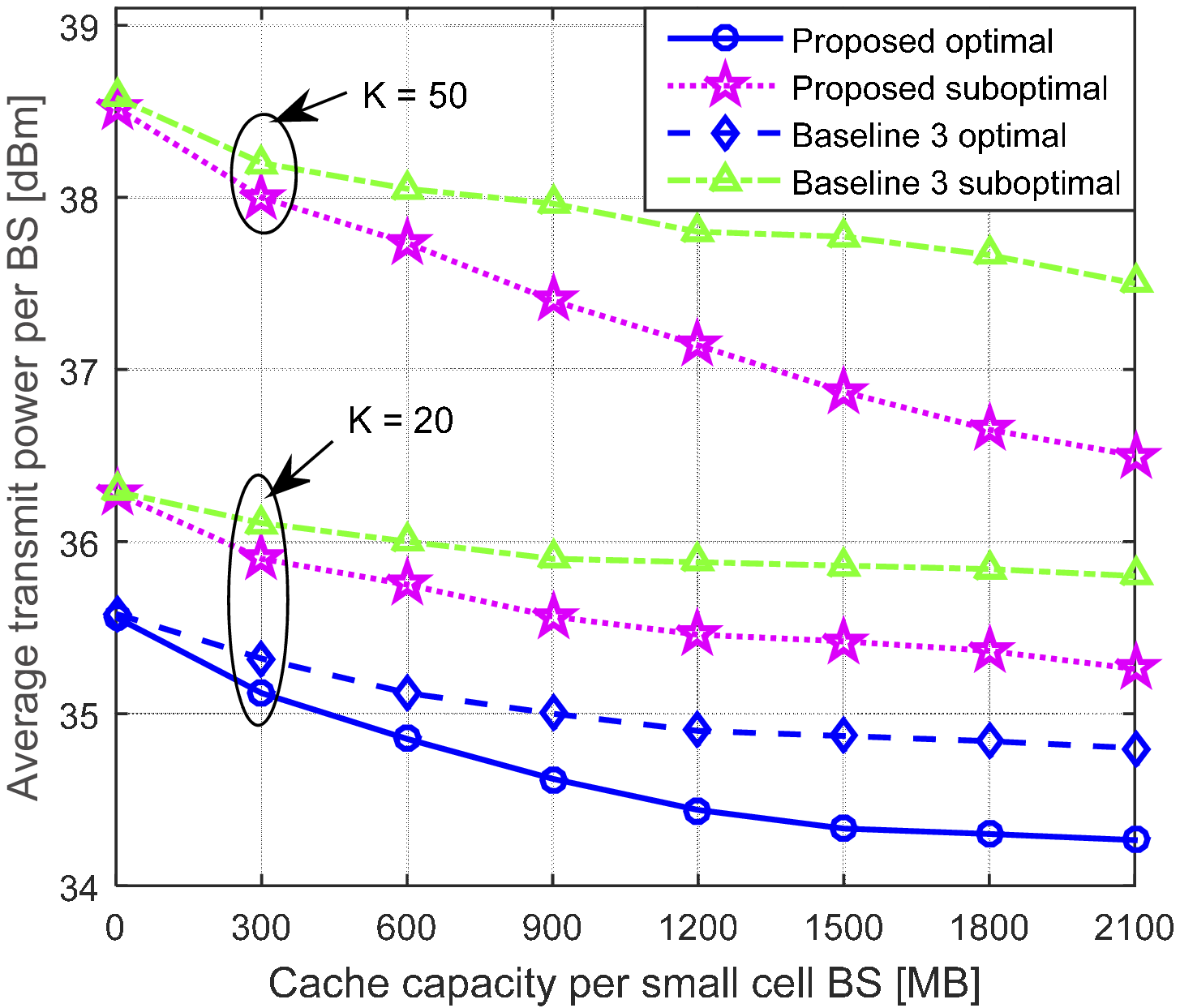} \label{fig6a}} \;
\subfloat[]{\includegraphics[width=3.0in]{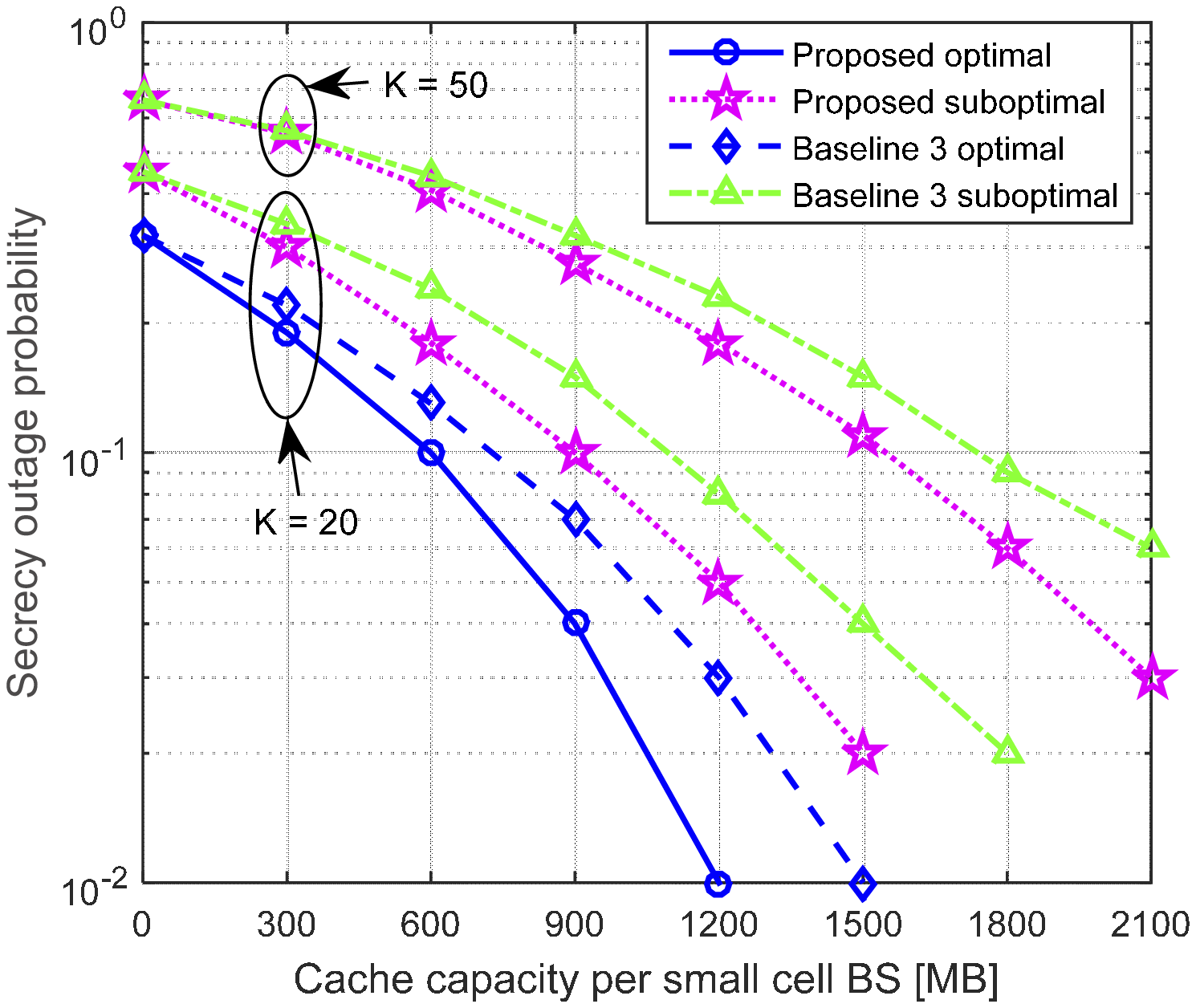} \label{fig6b}}
\protect\protect\caption{\small (a) Average transmit power per BS and (b) secrecy outage probability versus cache capacity for the proposed
optimal and suboptimal schemes and Baseline 3, where $M=K+1$ and $M_{u}=K/10$.}
\end{figure}

Finally, Figs.~\ref{fig6a} and~\ref{fig6b} show the performance of the proposed schemes and Baseline 3 for larger networks, where the number of users, $K$, the number of small cells, $M-1$, and the number of untrusted small cell BSs, $M_{u}$, satisfy $M=K+1$ and $M_{u}=K/10$. We consider a system bandwidth of 7~MHz to ensure that, despite the large number of users, the QoS requirements of each user can be fulfilled with high probability. For Baseline 3, the caching decisions are determined by employing Algorithms~\ref{alg1} and~\ref{alg2}, and the resulting schemes are referred to as ``Baseline 3 optimal'' and ``Baseline 3 suboptimal'', respectively. For $K=50$, only the performance of the proposed suboptimal and baseline suboptimal schemes is shown because of the high computational complexity of the optimal schemes. As $K$ increases, more untrusted BSs, $M_{u}=K/10$, are present in the network and each untrusted BS can eavesdrop a larger number of users. Hence, the available DoFs per BS for the delivery of the base-layer subfiles is reduced and the likelihood of data leakage is increased. Therefore, to ensure secure video streaming, for larger $K$, a larger average transmit power per BS is needed for AN transmission. On the other hand,  exploiting both trusted and untrusted BSs for cooperative transmission, the proposed schemes significantly outperform Baseline 3, particularly for networks with large numbers of users and large cache capacities.

\section{\label{sec:Conclusion}Conclusion}
In this paper, secure video streaming was investigated for small cell networks with untrusted small cell BSs which can intercept both cached and delivered video data. SVC coding and caching were jointly exploited to facilitate secure cooperative MIMO transmission and to not only mitigate the negative impact of the untrusted BSs but to exploit them for secrecy enhancement. A two-timescale non-convex robust optimization problem was formulated to optimize caching and delivery for minimization of the total BS transmit power required for secure video streaming with imperfect CSI knowledge. In the large timescale, the caching optimization problem was solved offline by a modified GBD algorithm. To reduce the computational complexity, a suboptimal caching algorithm was also studied. In the short timescale, the delivery optimization problem for a given cache status was solved online by SDP. Simulation results revealed that, compared to several baseline schemes, the proposed optimal and suboptimal schemes can significantly enhance both the secrecy and the power efficiency of video streaming in small cell networks as long as the total number of antennas at the trusted BSs exceeds that at the untrusted BSs. 

\appendix
\subsection{\label{proof1}Proof of Proposition~\ref{prop1}}

We begin the proof by defining an auxiliary optimization problem: 
\vspace{-.1cm}
\begin{align}
\textrm{\textrm{P2}:}\quad\minimize_{\mathbf{D}_{t},\,\mathbf{s}_{t},\,\mathbf{q}}\quad & \sum\nolimits _{t\in\mathcal{T}_{0}}\left[U_{\textrm{TP}}\left(\mathbf{q},\mathbf{D}_{t}\right)+f_{\textrm{Pen}}\left(\mathbf{s}_{t}\right)\right]\label{eq:aux}\\
\st\quad & \overline{\textrm{C3}}\textrm{, }\overline{\textrm{C8}}\textrm{, }\overline{\textrm{C9}}\textrm{, }\textrm{C12, }\mathbf{D}_{t}\in\mathcal{D}_{t},\;\mathbf{s}_{t}\succeq\mathbf{0},\;\mathbf{q}\in\mathcal{Q},\nonumber 
\end{align}
where $\mathcal{D}_{t}$ is given in \eqref{eq:Primal}. Problems P2 and \{\eqref{eq:Primal}, \eqref{eq:Master}\} are equivalent as $\min_{x\in\mathcal{X},\, y\in\mathcal{Y}}f\left(x,y\right) = \min_{y\in\mathcal{Y}}\left[\min_{x\in\mathcal{X}}f\left(x,y\right)\right]$~\cite{Boyd2004Convex}. 

Because of the perturbation, the feasible set of problem P2 is a superset of the feasible set of problem P1. Thus, if P1 is feasible, so is P2 (and \eqref{eq:Primal}). Moreover, the inequality constraint functions on the left hand sides of the big-M constraints  $\overline{\textrm{C3}}$, $\overline{\textrm{C8}}$, and $\overline{\textrm{C9}}$ are bounded from above, e.g., $\mathrm{tr} \! \left(\boldsymbol{ \Lambda}_{m} \mathbf{W}_{\boldsymbol{\rho},l,t} \right) \!-\! q_{f(\boldsymbol{\rho}),l,m} P_{m}^{\max} \!\le\! P_{m}^{\max}$ for $\overline{\textrm{C3}}$. Considering $\mathbf{s}_{t} \succeq \mathbf{0}$ on the right hand sides of $\overline{\textrm{C3}}$, $\overline{\textrm{C8}}$, and $\overline{\textrm{C9}}$, the feasibility statement in part i) of Proposition~\ref{prop1} thus always holds for any $\mathbf{q} \in \mathcal{Q}$.

Next, we show the optimality statement in part i) by contradiction. Assume that $\left( \mathbf{q}^{*},\mathbf{D}_{t}^{*}\right)$ solves P1. Then $\left(\mathbf{q}^{*},\mathbf{D}_{t}^{*},\mathbf{s}_{t} \right)$ with $\mathbf{s}_{t} \succeq \mathbf{0}$ is feasible for P2. Denote the objective function of P2 by $f\left( \mathbf{q},\mathbf{D}_{t}, \mathbf{s}_{t}\right)$. We have $f\left(\mathbf{q},\mathbf{D}_{t},\mathbf{0}\right) \ge f\left(\mathbf{q}^{*},\mathbf{D}_{t}^{*}, \mathbf{0} \right),  \forall (\mathbf{q}, \mathbf{D}_{t} )$. Besides, let $(\mathbf{q}^{+}, \mathbf{D}_{t}^{+}, \mathbf{s}_{t}^{+} )$ be the optimal solution of P2. If $(\mathbf{q}^{+}, \mathbf{D}_{t}^{+} ) \neq \left(\mathbf{q}^{*},\mathbf{D}_{t}^{*} \right)$, then $\mathbf{s}_{t}^{+}\neq\mathbf{0}$ necessarily holds. However, since $\mu\gg1$, we have $f_{\textrm{Pen}} \left(\mathbf{s}_{t}^{+} \right) > f\left(\mathbf{q}^{*},\mathbf{D}_{t}^{*},\mathbf{0}\right)$ and thus $f \left(\mathbf{q}^{+}, \mathbf{D}_{t}^{+}, \mathbf{s}_{t}^{+}\right) > f \left( \mathbf{q}^{*}, \mathbf{D}_{t}^{*}, \mathbf{0}\right)$, which contradicts the optimality of $\left(\mathbf{q}^{+}, \mathbf{D}_{t}^{+},\mathbf{s}_{t}^{+}\right)$. Therefore, part i) is proved.

Finally, we prove part ii). Obviously, i) implies that problem P1 is infeasible when $\nu \left(\mathbf{q}\right) = +\infty$. Assume that the optimal solution of problem P2, $\left( \mathbf{q}^{+}, \mathbf{D}_{t}^{+},\mathbf{s}_{t}^{+} \right)$, satisfies $\mathbf{s}_{t}^{+}\neq\mathbf{0}$. Then, a feasible solution of the form $\left( \mathbf{q}, \mathbf{D}_{t}, \mathbf{0} \right)$ does not exist for P2, since otherwise, $f\left( \mathbf{q}^{+}, \mathbf{D}_{t}^{+},\mathbf{s}_{t}^{+}\right) \le f\left( \mathbf{q}, \mathbf{D}_{t}, \mathbf{0} \right)$ has to hold. Therefore, P1 is also infeasible, which completes the proof.

\subsection{\label{proof2}Proof of Proposition \ref{prop2}}

As strong duality holds for \eqref{eq:Primal}, we know that $(\mathbf{D}_{t}^{k},\,\mathbf{s}_{t}^{k})$
minimizes the Lagrangian, i.e., $(\mathbf{D}_{t}^{k},\,\mathbf{s}_{t}^{k}) \in {\arg\min}_{\mathbf{D}_{t} \in \mathcal{D}_{t},\,\mathbf{s}_{t}\succeq\mathbf{0}}\mathcal{L}_{\mathbf{q}^{k}}(\mathbf{D}_{t},\mathbf{s}_{t}; \boldsymbol{\lambda}_{t}^{k})$. Since $\mathcal{L}_{\mathbf{q}^{k}}(\mathbf{D}_{t},\mathbf{s}_{t})=f_{1}(\mathbf{q}^{k};\boldsymbol{\lambda}_{t}^{k}) + f_{2}(\mathbf{D}_{t},\mathbf{s}_{t};\boldsymbol{\lambda}_{t}^{k})$, where $f_{1}(\mathbf{q}^{k};\boldsymbol{\lambda}_{t}^{k})$ is a constant, we also have $(\mathbf{D}_{t}^{k},\,\mathbf{s}_{t}^{k}) \in {\arg\min}_{\mathbf{D}_{t} \in \mathcal{D}_{t},\,\mathbf{s}_{t}\succeq\mathbf{0}}f_{2}(\mathbf{D}_{t},\mathbf{s}_{t}; \boldsymbol{\lambda}_{t}^{k})$. Finally, i) has to hold as $\mathcal{L}_{\mathbf{q}}\left(\mathbf{D}_{t},\mathbf{s}_{t}; \boldsymbol{\lambda}_{t}\right)$ is separable w.r.t. $\mathbf{q}$ and $\left\{ \mathbf{D}_{t},\mathbf{s}_{t}\right\} $ in \eqref{eq:RM}, cf. \eqref{eq:Lagrang}. 

Meanwhile, according to the Karush-Kuhn-Tucker (KKT) conditions for \eqref{eq:Primal}, we have
\vspace{-.1cm}
\begin{alignat*}{1}
\lambda_{\boldsymbol{\rho},l,m,t}^{\overline{\textrm{C3}},k}[\mathrm{tr}(\boldsymbol{ \Lambda}_{m} \mathbf{W}_{\boldsymbol{\rho},l,t}^{k}) - q_{f(\boldsymbol{\rho}),l,m}^{k}P_{m}^{\max}-s_{\boldsymbol{\rho},l,m,t}^{\textrm{C3},k}] & = 0, \\ 
\lambda_{\boldsymbol{\rho},l,j,t}^{\overline{\textrm{C8}},k}[\mathrm{tr}(\mathbf{W}_{\boldsymbol{\rho},l,t}^{k} - \overline{\mathbf{W}}_{\boldsymbol{\rho},l,j,t}^{k}) - q_{f(\boldsymbol{\rho}),l,j}^{k}P_{\max}-s_{\boldsymbol{\rho},l,j,t}^{\textrm{C8},k}] & =0,\\ 
\lambda_{\boldsymbol{\rho},l,j,t}^{\overline{\textrm{C9}},k}[\mathrm{tr}(\overline{\mathbf{W}}_{\boldsymbol{\rho},l,j,t}^{k})-(1-q_{f(\boldsymbol{\rho}),l,j}^{k})P_{\max}-s_{\boldsymbol{\rho},l,j,t}^{\textrm{C9},k}] & =0,  
\end{alignat*}
 Therefore, 
 \vspace{-.2cm}
\begin{alignat}{1}
& \min_{\mathbf{D}_{t}\in\mathcal{D}_{t},\,\mathbf{s}_{t}\succeq\mathbf{0}}f_{2}(\mathbf{D}_{t},\mathbf{s}_{t};\boldsymbol{\lambda}_{t}^{k}) \nonumber \\
&=f_{2}(\mathbf{D}_{t}^{k},\mathbf{s}_{t}^{k};\boldsymbol{\lambda}_{t}^{k}) \nonumber \\
&=U_{\mathrm{TP}}\left(\mathbf{q}^{k},\mathbf{D}_{t}^{k}\right)+f_{\textrm{Pen}}\left(\mathbf{s}_{t}^{k}\right)-f_{1}(\mathbf{q}^{k};\boldsymbol{\lambda}_{t}^{k}).\label{eq:psi2}
\end{alignat}
By substituting \eqref{eq:psi2} into $\mathcal{L}_{\mathbf{q}}(\mathbf{D}_{t},\mathbf{s}_{t}; \boldsymbol{\lambda}_{t}^{k})$,  \eqref{eq:psi} is established. This completes the proof. 

\subsection{\label{proof3}Proof of Proposition \ref{prop3}}

Let $\widehat{\mathbf{q}}\in\mathcal{Q}$ be a solution of \eqref{eq:RM}. By Algorithm \ref{alg1}, the optimality cut $\alpha\ge\xi(\mathbf{q};\widehat{\boldsymbol{\lambda}}_{t})$ is then generated for $\mathbf{q} = \widehat{\mathbf{q}}$. We have $\xi(\widehat{\mathbf{q}};\widehat{\boldsymbol{\lambda}}_{t})=\nu(\widehat{\mathbf{q}})$ due to the strong duality of the respective SDP subproblem \eqref{eq:Primal} for $\mathbf{q} = \widehat{\mathbf{q}}$. If $(\overline{\mathbf{q}}, \overline{\alpha})$ were to solve \eqref{eq:RM} again in another iteration with $\overline{\mathbf{q}} = \widehat{\mathbf{q}}$, then $\overline{\alpha}\ge\xi(\overline{\mathbf{q}};\widehat{ \boldsymbol{ \lambda}}_{t})$ and  $\overline{\alpha} \ge \nu(\widehat{\mathbf{q}}) = \nu(\overline{\mathbf{q}})$ would hold, which lead to the termination of Algorithm \ref{alg1} since $LB\ge UB$, cf. line~\ref{line2}. This implies that $\widehat{\mathbf{q}}$ does not repeat itself in intermediate iterations. Since set $\mathcal{Q}\subseteq\left\{ 0,1\right\} ^{F\times L\times M}$ is finite, Algorithm~\ref{alg1} has to converge in a finite number of iterations.

To prove the equivalence between P1 and P0, we show here that the solution of the relaxed problem P1 satisfies $\mathrm{rank}(\mathbf{W}_{\boldsymbol{\rho},l,t})=1$ with probability one. Let $\boldsymbol{\Upsilon}_{\boldsymbol{\rho},l,t}\succeq\mathbf{0}$,
$\boldsymbol{\Phi}_{\boldsymbol{\rho},l,t}\succeq\mathbf{0}$, and $\boldsymbol{ \Theta}_{\boldsymbol{\rho},l,t} \succeq \mathbf{0}$ be the Lagrange multipliers associated with constraints $\widetilde{\textrm{C6}}$, C10, and $\textrm{C12: } \mathbf{W}_{ \boldsymbol{\rho},l,t} \succeq \mathbf{0}$, respectively, where C12 is implied by C10. The Lagrangian of problem P1 is given by, 
\vspace{-.1cm}
\begin{alignat}{1}
&\mathcal{L}(\mathbf{W}_{\boldsymbol{\rho},l,t};\boldsymbol{\Upsilon}_{\boldsymbol{\rho},l,t}, \boldsymbol{\Phi}_{\boldsymbol{\rho},l,t},\boldsymbol{\Theta}_{\boldsymbol{\rho},l,t}) =  \nonumber \\
&\sum_{\boldsymbol{\rho},l}\mathrm{tr} \left[\left(\mathbf{B}{}_{\boldsymbol{\rho},l,t} - \boldsymbol{\Theta}_{\boldsymbol{\rho},l,t} - \tfrac{1}{\eta_{\boldsymbol{\rho},l}^{\textrm{req}}} \mathbf{U}_{\boldsymbol{\rho},t} \boldsymbol{\Upsilon}_{\boldsymbol{\rho},l,t}\mathbf{U}_{\boldsymbol{\rho},t}^{H}\right)\mathbf{W}_{\boldsymbol{\rho},l,t}\right] \!+\! \Delta_{2},\label{eq:app-36}
\end{alignat}
where $\mathbf{B}{}_{\boldsymbol{\rho},l,t} \triangleq \mathbf{I} + \boldsymbol{\Delta}_{1}-\boldsymbol{\Phi}_{\boldsymbol{\rho},l,t}$; and
$\boldsymbol{\Delta}_{1} \succeq \mathbf{0}$ and $\Delta_{2}\in\mathbb{R}$ denote the collection of terms that are relevant and irrelevant to $\mathbf{W}_{\boldsymbol{\rho},l,t}$, respectively. Hence, the dual problem of P1 is given by 
\vspace{-.1cm}
\begin{equation}
\max_{\boldsymbol{\Upsilon}_{\boldsymbol{\rho},l,t}\succeq\mathbf{0},\boldsymbol{\Phi}_{\boldsymbol{\rho},l,t}\succeq\mathbf{0},\boldsymbol{\Theta}_{\boldsymbol{\rho},l,t}\succeq\mathbf{0}}\; \min_{\mathbf{W}_{\boldsymbol{\rho},l,t}}  \mathcal{L}(\mathbf{W}_{\boldsymbol{\rho},l,t};\boldsymbol{\Upsilon}_{\boldsymbol{\rho},l,t},\boldsymbol{\Phi}_{\boldsymbol{\rho},l,t},\boldsymbol{\Theta}_{\boldsymbol{\rho},l,t}).\label{eq:kkt3}
\end{equation}

We now define 
\begin{equation}
\boldsymbol{\Upsilon}_{\boldsymbol{\rho},l,t}\triangleq\Bigg[\begin{array}{cc}
\overline{\boldsymbol{\Upsilon}}_{\boldsymbol{\rho},l,t} & \overline{\boldsymbol{\gamma}}_{\boldsymbol{\rho},l,t}\\
\overline{\boldsymbol{\gamma}}_{\boldsymbol{\rho},l,t}^{H} & \alpha_{\boldsymbol{\rho},l,t}
\end{array}\Bigg]\in\mathbb{C}^{(N+1)\times(N+1)},\label{eq:app-37}
\end{equation}
where $\overline{\boldsymbol{\Upsilon}}_{\boldsymbol{\rho},l,t} \succeq \mathbf{0}$, $\alpha_{\boldsymbol{\rho},l,t}\ge0$ and $\overline{\boldsymbol{\gamma}}_{\boldsymbol{\rho},l,t}$ is chosen to ensure $\boldsymbol{\Upsilon}_{\boldsymbol{\rho},l,t}\succeq\mathbf{0}$. If P0 is feasible, so is P1, and then $\mathbf{W}_{ \boldsymbol{\rho},l,t} \neq \mathbf{0}$. Moreover, as strong duality holds for SDP subproblem \eqref{eq:Primal}, the optimal beamformers and the optimal dual solutions satisfy the  KKT optimality conditions. In particular, by substituting $\mathbf{U}_{\boldsymbol{\rho},t}=[\mathbf{I}_{N},\widehat{ \mathbf{h}}_{\boldsymbol{\rho},t}]$ and \eqref{eq:app-37} into \eqref{eq:app-36}, we have  
\vspace{-.1cm}
\begin{alignat}{1}
\overline{\mathbf{B}}{}_{\boldsymbol{\rho},l,t}-\tfrac{\alpha_{\boldsymbol{\rho},l,t}}{\eta_{\boldsymbol{\rho},l}^{\textrm{req}}}\mathbf{\widehat{h}}_{\boldsymbol{\rho},t}\mathbf{\widehat{h}}_{\boldsymbol{\rho},t}^{H} & =\boldsymbol{\Theta}_{\boldsymbol{\rho},l,t},\label{eq:kkt1} \\
\boldsymbol{\Theta}_{\boldsymbol{\rho},l,t}\mathbf{W}_{\boldsymbol{\rho},l,t} & =\mathbf{0},\label{eq:kkt2}
\end{alignat}
where $\overline{\mathbf{B}}{}_{\boldsymbol{\rho},l,t} = \mathbf{B}{}_{\boldsymbol{\rho},l,t} - \tfrac{1}{\eta_{\boldsymbol{\rho},l}^{\textrm{req}}} \left( \overline{\boldsymbol{\Upsilon}}_{\boldsymbol{\rho},l,t}+\overline{\boldsymbol{\gamma}}_{\boldsymbol{\rho},l,t} \widehat{\mathbf{h}}_{\boldsymbol{\rho},t}^{H}+\overline{\boldsymbol{\gamma}}_{\boldsymbol{\rho},l,t}^{H}\widehat{\mathbf{h}}_{\boldsymbol{\rho},t}\right)$.

Next, we show by contradiction that $\overline{\mathbf{B}}{}_{\boldsymbol{\rho},l,t} \succ \mathbf{0}$ holds with probability one. Assume that $\overline{ \mathbf{B}}{}_{\boldsymbol{\rho},l,t}$ has at least one non-positive eigenvalue $\tau\le0$ and the corresponding eigenvector is $\widetilde{ \mathbf{w}}_{\boldsymbol{\rho},l,t}$,
i.e., $(\overline{\mathbf{B}}{}_{\boldsymbol{\rho},l,t} - \tau\mathbf{I}) \widetilde{\mathbf{w}}_{\boldsymbol{\rho},l,t}=\mathbf{0}$.
Let $\mathbf{W}_{\boldsymbol{\rho},l,t} = \beta \widetilde{\mathbf{w}}_{\boldsymbol{\rho},l,t} \widetilde{\mathbf{w}}_{\boldsymbol{\rho},l,t}^{H} \succeq \mathbf{0}$,
where $\beta>0$. By substituting $\mathbf{W}_{\boldsymbol{\rho},l,t}$ into \eqref{eq:kkt3}, we further have 
\vspace{-.1cm}
\begin{alignat}{1}
& \mathcal{L}(\mathbf{W}_{\boldsymbol{\rho},l,t};\boldsymbol{\Upsilon}_{\boldsymbol{\rho},l,t},\boldsymbol{\Phi}_{\boldsymbol{\rho},l,t},\boldsymbol{\Theta}_{\boldsymbol{\rho},l,t}) = \underset{^{\le0}}{\beta\underbrace{\tau\sum_{\boldsymbol{\rho},l}\widetilde{\mathbf{w}}_{\boldsymbol{\rho},l,t}^{H}\widetilde{\mathbf{w}}_{\boldsymbol{\rho},l,t}}} \nonumber \\
&\qquad\qquad\qquad  -\beta\sum_{\boldsymbol{\rho},l}\tfrac{\alpha_{\boldsymbol{\rho},l,t}}{\eta_{\boldsymbol{\rho},l}^{\textrm{req}}}\mathbf{\widehat{h}}_{\boldsymbol{\rho},t}^{H}\widetilde{\mathbf{w}}_{\boldsymbol{\rho},l,t}\widetilde{\mathbf{w}}_{\boldsymbol{\rho},l,t}^{H}\mathbf{\widehat{h}}_{\boldsymbol{\rho},t} \!+\! \Delta_{2}.
\end{alignat}
Hence, if $\alpha_{\boldsymbol{\rho},l,t} = 0$, then $\widetilde{ \mathbf{w}}_{\boldsymbol{\rho},l,t} = \mathbf{0}$ necessarily holds to ensure $\mathcal{L} > -\infty$; yet this contradicts the condition that $\mathbf{W}_{\boldsymbol{\rho},l,t}\neq\mathbf{0}$.
On the other hand, if $\alpha_{\boldsymbol{\rho},l,t}>0$, then the minimum value of \eqref{eq:kkt3} is obtained for $\beta \to \infty$, since $\mathbf{h}_{ \boldsymbol{\rho},t}$ is statistically independent and $- \beta \tfrac{ \alpha_{\boldsymbol{\rho},l,t}}{\eta_{\boldsymbol{\rho},l}^{\textrm{req}}} \mathbf{ \widehat{h}}_{\boldsymbol{\rho},t}^{H} \widetilde{ \mathbf{w}}_{\boldsymbol{\rho},l,t} \widetilde{\mathbf{w}}_{\boldsymbol{\rho},l,t}^{H} \mathbf{ \widehat{h}}_{\boldsymbol{\rho},t} \to - \infty$ with probability one. That is, the dual problem \eqref{eq:kkt3} is unbounded from below, and consequently, the primal problem is infeasible, which is also a contradiction. Therefore, $\overline{\mathbf{B}}{}_{\boldsymbol{\rho},l,t} \succ \mathbf{0}$ is proved.

Finally, based on \eqref{eq:kkt1}, \eqref{eq:kkt2}, and $\overline{ \mathbf{B}}_{ \boldsymbol{\rho},l,t} \succ \mathbf{0}$, we have, 
\vspace{-.2cm}
\begin{align}
\mathrm{rank}(\mathbf{W}_{\boldsymbol{\rho},l,t}) & \overset{\textrm{(a)}}{=}\mathrm{rank}(\overline{\mathbf{B}}{}_{\boldsymbol{\rho},l,t}\mathbf{W}_{\boldsymbol{\rho},l,t}) \nonumber \\
 &\overset{\textrm{(b)}}{=}\mathrm{rank} \big(\tfrac{\alpha_{\boldsymbol{\rho},l,t}}{\eta_{\boldsymbol{\rho},l}^{\textrm{req}}}\mathbf{W}_{\boldsymbol{\rho},l,t} \mathbf{\widehat{h}}_{\boldsymbol{\rho},t} \mathbf{\widehat{h}}_{\boldsymbol{\rho},t}^{H} \big) \nonumber \\
 & \overset{\textrm{(c)}}{\le} \min \left\{ \mathrm{rank} \big(\tfrac{\alpha_{\boldsymbol{\rho},l,t}}{\eta_{\boldsymbol{\rho},l}^{\textrm{req}}}\mathbf{W}_{\boldsymbol{\rho},l,t}\big),\mathrm{rank}\big(\mathbf{\widehat{h}}_{\boldsymbol{\rho},t}\mathbf{\widehat{h}}_{\boldsymbol{\rho},t}^{H}\big)\right\} \nonumber \\
 &\le1,
\end{align}
where (a) is due to $\overline{\mathbf{B}}{}_{\boldsymbol{\rho},l,t}\succ\mathbf{0}$,
(b) is a result of \eqref{eq:kkt1} and \eqref{eq:kkt2}, and (c) follows from the basic rank inequality $\mathrm{rank}(\mathbf{A} \mathbf{B})\le\min\left\{ \mathrm{rank}(\mathbf{A}),\mathrm{rank}(\mathbf{B})\right\} $. On the other hand, since $\mathbf{W}_{ \boldsymbol{\rho},l,t} \neq \mathbf{0}$, the condition $\mathrm{rank}(\mathbf{W}_{\boldsymbol{\rho},l,t})=1$ holds with probability one. This completes the proof.

\end{document}